\pdfoutput=1

\documentclass{jpp}
\usepackage{graphicx}

\usepackage[utf8]{inputenc}
\usepackage[T1]{fontenc}
\usepackage{amsmath}
\usepackage{subcaption}
\usepackage{braket}
\usepackage{bbold}
\usepackage{graphicx} 
\usepackage{xcolor}
\usepackage[normalem]{ulem}

\shorttitle{Numerical Simulations of 3D Ion Crystal Dynamics}
\shortauthor{J. Zaris, W. Johnson, A. Shankar, J.J. Bollinger, and S.E. Parker}

\title{Numerical Simulations of 3D Ion Crystal Dynamics in a Penning Trap using the Fast Multipole Method}

\author{John Zaris\aff{1}
  \corresp{\email{john.zaris@colorado.edu}},
  Wes Johnson\aff{1},
  Athreya Shankar\aff{2},
  John J. Bollinger\aff{3},
 \and Scott E. Parker\aff{1,4}}

\affiliation{\aff{1}Department of Physics, University of Colorado, Boulder, CO 80309, USA
\aff{2}Department of Instrumentation and Applied Physics, Indian Institute of Science, Bangalore, India, 560012
\aff{3}National Institute of Standards and Technology, Boulder, Colorado 80309, USA
\aff{4}Renewable and Sustainable Energy Institute, University of Colorado, Boulder, Colorado 80309, USA}

\begin{document}

\maketitle

\begin{abstract}
We simulate the dynamics, including laser cooling, of 3D ion crystals confined in a Penning trap using a newly developed molecular dynamics-like code.  The numerical integration of the ions' equations of motion is accelerated using the fast multipole method to calculate the Coulomb interaction between ions, which allows us to efficiently study large ion crystals with thousands of ions.  In particular, we show that the simulation time scales linearly with ion number, rather than with the square of the ion number.  By treating the ions' absorption of photons as a Poisson process, we simulate individual photon scattering events to study laser cooling of 3D ellipsoidal ion crystals.  Initial simulations suggest that these crystals can be efficiently cooled to ultracold temperatures, aided by the mixing of the easily cooled axial motional modes with the low frequency planar modes.  In our simulations of a spherical crystal of 1,000 ions, the planar kinetic energy is cooled to several millikelvin in a few milliseconds while the axial kinetic energy and total potential energy are cooled even further. This suggests that 3D ion crystals could be well-suited as platforms for future quantum science experiments.
\end{abstract}

\section{Introduction}
Studies of ion crystals confined in Penning traps have produced important results and applications in the fields of atomic physics \citep{wineland1983laser,arnold2015prospects}, quantum information science \citep{bohnet2016quantum,garttner2017measuring,gilmore2021quantum}, nonneutral plasma physics \citep{dubin1999trapped,dubin2020normal}, and high energy physics \citep{budker2022millicharged}.  In a variety of instances, numerical simulations of such crystals have been used to investigate their classical degrees of freedom and have proven useful in guiding experimental protocols.  For instance, crystal equilibria have long been studied in simulations employing either full or guiding-center approximated dynamics \citep{dubin1988computer,hasse1991structure}.  More recently, simulations of ion crystals have been utilized to study the resolution of motional modes \citep{shankar2020broadening} and to improve the laser cooling schemes used to cool down ion crystals for quantum science experiments \citep{johnson2023rapid}.

There are a few characteristics of Penning trap ion crystals which account for their use in such a variety of fields.  First, their ultracold temperatures make them great candidates for studies in which precise quantum control is needed. For instance, ion spin states of a crystal have been entangled with its ultracold motional modes, allowing for the detection of displacements below the standard quantum limit \citep{gilmore2021quantum}.  Secondly, since the ion crystals consist of charged particles, it is possible, for example, to engineer ion-ion interactions \citep{britton2012engineered, bohnet2016quantum} or to search for interactions with charged particles beyond the standard model \citep{budker2022millicharged}. Furthermore, large numbers of ions are routinely trapped, leading to excellent experimental sensitivity in quantum sensing protocols which, in the future, may be applied to detect dark matter \citep{affolter2020phase, gilmore2021quantum}. 

For these reasons, it is advantageous to develop a numerical simulation which can study cold crystals with large numbers of ions.  Using previous simulation frameworks, the dynamics and laser cooling of small planar crystals with tens to hundreds of ions have been studied \citep{tang2019first,johnson2023rapid}.  We aim to build on this work by studying larger 3D crystals with thousands of ions.  In previous studies, it has proven difficult to simulate the dynamics of large numbers of ions, as the number of Coulomb interaction calculations scales with the square of the ion number.  In order to avert this prohibitive scaling, we implement the fast multipole method (FMM), which uses multipole expansions of collections of source charges to approximate the electrostatic potential at the location of a sufficiently distant target charge \citep{greengard1987fast,greengard1997new}.  For large ion numbers, the simulation time scales linearly with ion number (rather than its square) when the FMM is used.  This allows us to accurately and efficiently simulate the dynamics of large crystals.

In a Penning trap, ions are confined axially by employing an electrostatic potential.  Since this potential necessarily is repulsive in the planar directions via Gauss's Law, an axial magnetic field is introduced to confine the ions in the plane.  In particular, the guiding center of an ion's trajectory experiences an $\boldsymbol{E}\times\boldsymbol{B}$ drift and undergoes circular motion about the symmetry axis of the trap.  In our simulations, we specifically model the NIST Penning trap \citep{gilmore2021quantum} which, in addition to the aforementioned electrostatic potential, includes a time-dependent planar electric potential, known as a rotating wall, to stabilize the crystal's rotation frequency about the symmetry axis \citep{huang1998precise}.

Understanding the motional modes of trapped ion crystals is critical when implementing effective Doppler laser cooling protocols.  Previous theoretical and numerical works have studied the modes of ultracold 2D planar crystals, in which the equilibrium axial coordinate of each ion is $z=0$ and the crystal's axial and planar motional modes decouple \citep{wang2013phonon,shankar2020broadening}.  In particular, it has been shown in simulations that the so called $\boldsymbol{E}\times \boldsymbol{B}$ planar modes are difficult to cool due to their large potential energy component \citep{johnson2023rapid}.  In this work, we simulate the laser cooling of 3D crystals, in which the motional modes are no longer purely axial or planar.  We find that the analogous $\boldsymbol{E}\times \boldsymbol{B}$ modes in these crystals can likely be cooled to $<1$ mK with relative ease.  We also study the cooling of the kinetic energy, both numerically and theoretically.  We show that the kinetic energy of 3D crystals can be efficiently cooled to the mK regime and obtain agreement between numerical and theoretical cooling predictions.

\section{\label{sec:ion_dynamics}Penning trap simulation model}

\subsection{\label{sec:theory}Penning trap physics}

In a Penning trap, ions are confined using static electric and magnetic fields.  An electric field, created by applying voltages to the trap's endcap electrodes, provides a confining force in the axial ($\hat{\boldsymbol{z}}$) direction but a repulsive force in the planar directions.  Near the center of the trap, at position $\boldsymbol{x} = x\hat{\boldsymbol{x}}+y\hat{\boldsymbol{y}}+z\hat{\boldsymbol{z}}$, the corresponding electrostatic potential is approximately harmonic and is parameterized by its strength $k_z$:

\begin{equation}
\label{pot_trap}
    \phi_{trap}(\boldsymbol{x}) = \frac{1}{4}k_z(2z^2-x^2-y^2).
\end{equation}

The strength $k_z$ is related to the axial frequency $\omega_z = \sqrt{qk_z/m}$, where $m$ is the ion mass and $q$ is its charge.  An axial magnetic field $\boldsymbol{B} = B\hat{\boldsymbol{z}}$ provides radial confinement.  The final external force on the ions is provided by the rotating wall potential, which has strength parameterized by $\delta$ and is important experimentally as it stabilizes the rotation frequency of the crystal at $\omega_r$. It is given by 

\begin{equation}
\label{pot_wall}
    \phi_{wall}(\boldsymbol{x},t) = \frac{1}{4}k_z\delta(x^2+y^2)\cos[2(\varphi+\omega_r t)].
\end{equation}

In the above definition, $\varphi$ is the azimuthal angle of the ion within the trap. Defining $\boldsymbol{A} = -yB\hat{\boldsymbol{x}}$. such that $\boldsymbol{B} = \boldsymbol{\nabla}\times\boldsymbol{A}$, the Hamiltonian of an $N$ ion crystal is then given by

\begin{equation}
\label{hamiltonian}
\begin{split}
    H &= \sum_{i=1}^N \frac{(\boldsymbol{p}_i-q_i\boldsymbol{A}(\boldsymbol{x}_i))^2}{2m_i}+\sum_{i=1}^N\Big[q_i\Big(\phi_{trap}(\boldsymbol{x}_i)+\phi_{wall}(\boldsymbol{x}_i,t)+\frac{1}{8\pi\epsilon_0}\sum_{j=1,j\neq i}^N\frac{q_j}{|\boldsymbol{x}_i-\boldsymbol{x}_j|}\Big)\Big]\\
    &=H_0+\sum_{i=1}^Nq_i\phi(\boldsymbol{x}_i).
\end{split}
\end{equation}

While our simulation of the ion crystal dynamics is carried out in the lab frame, it is often useful to transform the ion coordinates to the frame which rotates at frequency $\omega_r$, in which the explicit time dependence of the Hamiltonian vanishes. The transformation to the rotating frame is given by

\begin{equation}
\label{rotation}
\begin{split}
    \begin{pmatrix} x_r\\y_r\end{pmatrix} = \begin{pmatrix} \cos(\omega_r t) & -\sin(\omega_r t)\\ \sin(\omega_r t) & \cos(\omega_r t)\end{pmatrix}\begin{pmatrix} x\\y\end{pmatrix}.
\end{split}
\end{equation}

Introducing the cyclotron frequency, $\omega_c=qB/m$, the potential energy in this frame can be expressed as

\begin{equation}
\label{eq:pe_rot}
\begin{split}
    q_i\phi_{r}(\boldsymbol{x}_{r,i}) &= \frac{1}{2}m\omega_z^2z_{r,i}^2-\frac{1}{2}m\Big(\omega_r^2-\omega_c\omega_r+\frac{1}{2}\omega_z^2\Big)(x_{r,i}^2+y_{r,i}^2)+\frac{1}{4}qk_z\delta(x_{r,i}^2-y_{r,i}^2) \\
    &\;\;\;\;\;\;\;\;\;\;\;\;\;\;\;\;\;\;\;\;+\frac{q_i}{8\pi\epsilon_0}\sum_{j\neq i}\frac{q_j}{|\boldsymbol{x}_i-\boldsymbol{x}_j|}\\
    & = \frac{1}{2}m\omega_z^2\Big[C_xx_{r,i}^2+C_yy_{r,i}^2+C_zz_{r,i}^2\Big]+\frac{q_i}{8\pi\epsilon_0}\sum_{j\neq i}\frac{q_j}{|\boldsymbol{x}_i-\boldsymbol{x}_j|}.\\
\end{split}
\end{equation}

Here, $C_x=\beta-\delta$, $C_y=\beta+\delta$, and $C_z=1$, where $\beta$ is the ratio of the planar and axial confining potentials given by

\begin{equation}
\label{beta}
\beta = \frac{\omega_r(\omega_c-\omega_r)}{\omega_z^2}-\frac{1}{2}.
\end{equation}

Ion crystals confined in Penning traps are often laser cooled to millikelvin temperatures. In the NIST trap, $^9Be^+$ ions are cooled via the $2s^2S_{1/2}\rightarrow 1p^2P_{3/2}$ laser cooling transition.  The laser cooling setup considered for 3D crystals is shown in figure 1, and is quite similar to that which has been used to cool planar crystals in past experiments \citep{bohnet2016quantum, gilmore2021quantum}.  Two axial laser beams, detuned from the cooling transition resonance by $\Delta_{\parallel}$, provide cooling along the trap's symmetry axis.  These laser beams have relatively large widths and their intensity is, in fact, nearly uniform over the extent of the crystal.  Further cooling is provided by a planar beam which is detuned from the cooling transition resonance by $\Delta_{\perp}$ and whose wavevector is parallel to the $\hat{x}$ axis, but offset by a distance $d$. Ions near the beam center are moving in the direction of the beam due to the global rotation of the ion crystal.

\subsection{Simulation methodology}

Our simulation code uses a cyclotronic integrator \citep{patacchini2009explicit} to advance ion positions and velocities.  This scheme is identical to that described in \citet{tang2019first}. The time evolution operator $U(t,\Delta t)$ which advances positions $\boldsymbol{x}_i = (x_i,y_i,z_i)$ and $\boldsymbol{v}_i= (v_{x_i},v_{y_i},v_{z_i})$ from time $t$ to time $t+\Delta t$ satisfies the following relations:

\begin{equation}
\label{time_ev1}
\begin{pmatrix}\boldsymbol{x}_i(t+\Delta t)\\ \boldsymbol{v}_i(t+\Delta t)\end{pmatrix}=U(t,\Delta t)\begin{pmatrix}\boldsymbol{x}_i(t)\\ \boldsymbol{v}_i(t)\end{pmatrix},
\end{equation}

\begin{equation}
\label{time_ev2}
U(t,\Delta t)=U_0(\Delta t / 2) U_{kick}(t+\Delta t / 2 ; \Delta t) U_0(\Delta t / 2).
\end{equation}

Here $U_0$ evolves the position and velocity of each ion according to $H_0$.  In a static, uniform magnetic field, the rotation of a particle's trajectory can be written in a simple, analytical form, which determines the form of $U_0$. Explicitly, $U_0$ is written as

\begin{equation}
\label{U0}
U_0(\Delta t) = \begin{pmatrix}
    1&0&0&\sin(\omega_c\Delta t)/\omega_c&[\cos(\omega_c\Delta t)-1]/\omega_c&0\\0&1&0&-[\cos(\omega_c\Delta t)-1]/\omega_c&\sin(\omega_c\Delta t)/\omega_c&0\\0&0&1&0&0&\Delta t\\0&0&0&\cos(\omega_c\Delta t)&-\sin(\omega_c\Delta t)&0\\0&0&0&\sin(\omega_c\Delta t)&\cos(\omega_c\Delta t)&0\\0&0&0&0&0&0
\end{pmatrix}.
\end{equation}

$U_{kick}$ updates the ion velocities in response to all forces except that exerted by the magnetic field. The first argument of $U_{kick}$ in equation \ref{time_ev2} signifies that these forces are evaluated at the midpoint of the time interval $(t,t+\Delta t)$ (recall that $\phi_{wall}$ is time-dependent). The second argument implies that the operator is applied for time $\Delta t$, the total duration of the interval. In addition to accounting for the electrostatic potential $\phi$, $U_{kick}$ includes laser cooling effects via a simple model which computes the number of photons that each ion absorbs during the timestep.  Laser cooling is described in more detail in appendix \ref{appB}.  The theoretical rate at which an ion absorbs photons from laser beam $l$ is given by 

\begin{equation}
\label{scattering_rate}
 \gamma_L^l(\boldsymbol{x},\boldsymbol{v}) =  S_l(\boldsymbol{x})\gamma_0\frac{(\gamma_0/2)^2}{(\gamma_0/2)^2(1+2S_l(\boldsymbol{x}))+(\Delta_l-\boldsymbol{k}_l\cdot\boldsymbol{v})^2},
\end{equation}
where $\gamma_0 = 2\pi\times 18$ MHz is the natural linewidth of the laser-cooling transition, $\Delta_l$ is the detuning of the laser beam from the cooling transition, $\boldsymbol{k}_l$ is the beam's wavevector, and $S_l$ is the beam saturation intensity, which is proportional to its intensity.  In this paper, we study laser beams with 2D Gaussian intensity profiles.  The number of photons from laser $l$ scattered by ion $i$ between times $t$ and $t+\Delta t$, denoted $n_{i,l}(t)$, is found by generating a random integer from a Poisson distribution with mean $\Bar{n}_{i,l} = \gamma_L^l(\boldsymbol{x}_i,\boldsymbol{v}_i)\Delta t$, where $\boldsymbol{x}_i$ and $\boldsymbol{v}_i$ are evaluated at the midpoint of the timestep, in accordance with equation \ref{time_ev2}.  The change in the ion's velocity due to all scattered photons during the timestep is then computed as 

\begin{equation}
\label{laser_kick}
\Delta v_i^{lasers} = \sum_{l}\frac{n_{i,l}(t)\hbar\boldsymbol{k}_l}{m} + \sum_{l}\sum_{j=1}^{n_{i,l}}\frac{\hbar k_l\hat{\boldsymbol{q}}_{l,j}}{m}.
\end{equation}

Here, $\hat{{\boldsymbol{q}}}_{l,j}$ is a random unit vector representing the direction in which a given absorbed photon is emitted. The simulation timestep and saturation parameters of the laser beams are sufficiently small that each ion usually absorbs either zero or one photon per timestep.  Combining the above considerations, the operator $U_{kick}(t+\Delta t/2;\Delta t)$ updates the ion velocities according to 

\begin{equation}
\label{velocity_kick}
\Delta \boldsymbol{v}_i = -\frac{q\nabla\phi(\boldsymbol{x}_i)|_{t+\Delta t/2}\Delta t}{m} + \Delta v_i^{lasers}.
\end{equation}

The approach described here is similar to the Boris integrator commonly used in computational plasma physics \citep{birdsall2018plasma}. However, while $U_0$ advances both the positions and velocities, the analogous operator in the Boris scheme advances only the positions and, in doing so, does not account for the magnetic field. Instead, the effect of the magnetic field is included in the $U_{kick}$ operator, which itself advances only the velocities. Unlike the Boris integrator, the cyclotronic integrator is symplectic for a uniform magnetic field, so physical invariants of the system such as its energy remain nearly constant. 

\begin{figure}
    \centering
    \includegraphics[scale=0.3]{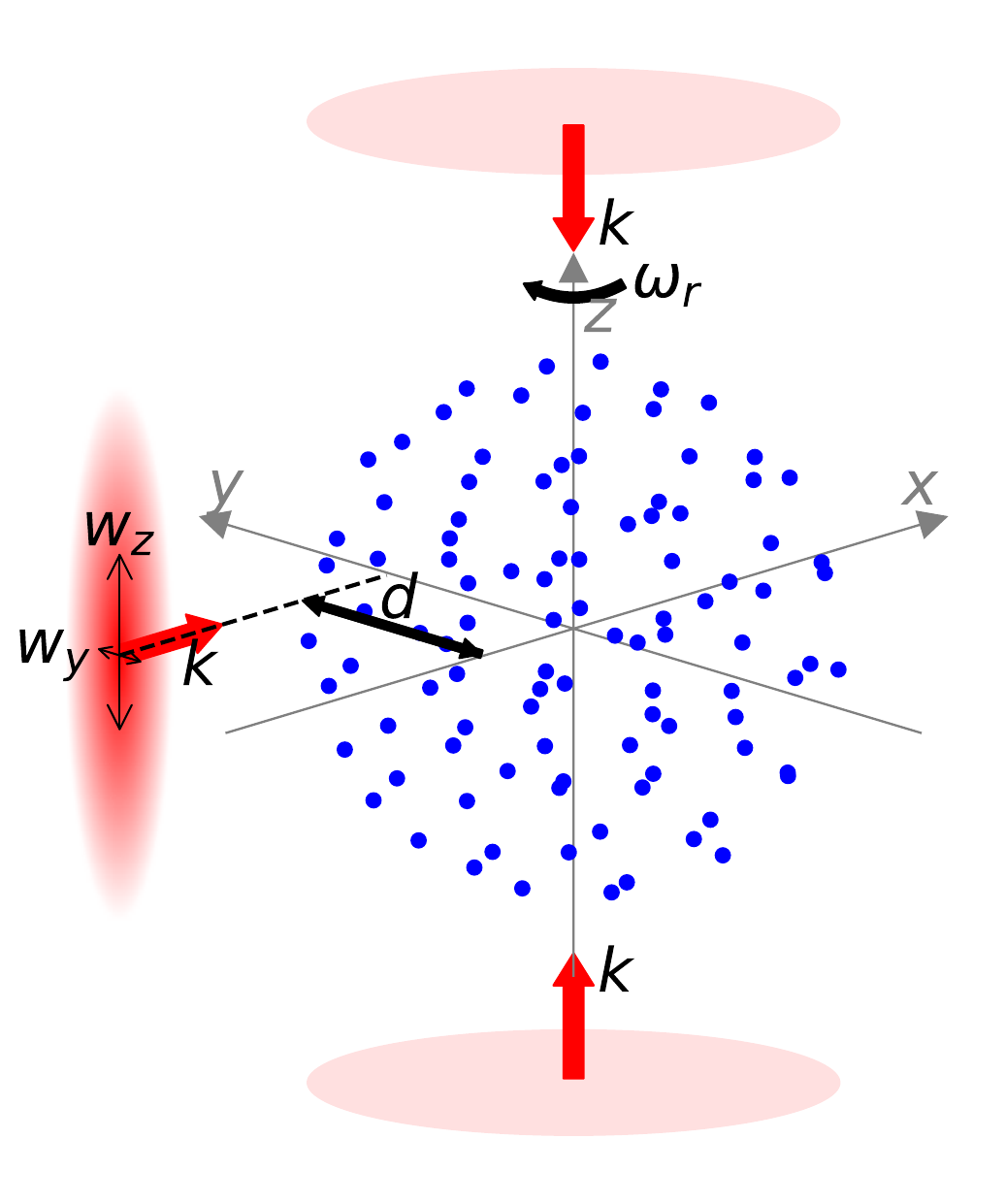}
    ~ 
    \caption{Proposed laser cooling setup for a 3D ion crystal confined in a Penning trap.  The setup is similar to that which has been implemented to cool planar crystals.  Two axial beams, detuned  from the $2s^2S_{1/2}\rightarrow 1p^2P_{3/2}$ laser cooling transition by $\Delta_{\parallel} = -\gamma_0/2$ (where $\gamma_0=2\pi\times 18$ MHz is the natural linewidth of the transition) cool the axial motion of the crystal.  These beams have large waists and their intensities are assumed to be uniform over the extent of the crystal.  In the NIST experiment, only one axial beam is used.  Two beams are used in our numerical simulations in order to prevent exciting the axial center of mass mode when the beams are turned off.  A planar beam, directed parallel to the $x$ axis and offset from it by a distance $d$, provides further cooling in the plane.  This beam is characterized by detuning $\Delta_{\perp}$ and beam waists $w_y$ and $w_z$ in the $y$ and $z$ directions, respectively.  In our simulations, we assume $w_z$ is large relative to the axial extent of the crystal in order to maximize cooling.}
    \label{cooling_setup}
\end{figure}

\subsection{Fast multipole method}
\label{sec: fmm}

As the number of simulated ions increases, the calculation of the Coulomb force on each ion becomes prohibitively slow.  While the time required for all other force calculations scales as $O(N)$, the time for the Coulomb calculation scales as $O(N^2)$, since the force on a given ion depends on the positions of all other ions.  To solve this issue, we utilize the fast multipole method, an algorithm which calculates the Coulomb force in a time which scales like $O(N)$ for large $N$.  In particular, our code uses the FMM3D library, which uses shared-memory parallelism to further reduce computation time \citep{greengard1997new,greengard1998accelerating,cheng1999fast,greengard2002new}.  In this section we provide a brief but complete overview of the FMM.

\subsubsection{Representations of electric potential}

The version of the FMM used in the FMM3D library is the result of decades of research aimed at improving computational efficiency.  We will provide a brief qualitative review of the FMM methodology, neglecting many mathematical details and algorithmic subtleties.  For a more complete description of the FMM algorithms important to this work, see \cite{greengard1997new,cheng1999fast}.  The central idea behind the FMM is that the Coulomb potential due to a collection of ions can be approximated by its multipole expansion (ME).  In turn, the potential at the location of a given ion sourced by an arbitrary collection of other ions can often be found quickly (in comparison to the direct calculation) by approximating the potential due to distant ions using their MEs.  Consider $n$ ions of charge $q$ located at spherical coordinates $(\rho_i,\theta_i,\varphi_i)$, such that all ions are located within a cube of side length $a$ centered on the origin.   Then at any point $(\rho,\theta,\varphi)$ with $\rho>\sqrt{3/4}a$, the Coulomb potential $\phi_c$ is given by

\begin{subequations}
\label{FMM_coul}
\begin{align}
\phi_c(\rho,\theta,\phi) = \sum_{n=0}^{\infty}\sum_{m=-n}^{n}\frac{M_n^m}{\rho^{n+1}}Y_n^m(\theta,\varphi)\label{FMM_coul1},\\    
M_n^m \equiv q\sum_{k=1}^n\rho_k^nY_n^{-m}(\theta_k,\varphi_k)\label{FMM_coul2}.
\end{align}
\end{subequations}

This is equivalent to theorem 2.1 in \cite{cheng1999fast} except that the source charges are located in a cube rather than a sphere.  This is a ME centered on the origin and $Y_n^m$ is the spherical harmonic of degree $n$ and order $m$. In practice, only a finite number of terms, $p$, are calculated, and the resulting truncation error is bounded by

\begin{equation}
\label{multipole_error}
\Big|\phi_c(\rho,\theta,\varphi) - \sum_{n=0}^{p}\sum_{m=-n}^{n}\frac{M_n^m}{\rho^{n+1}}Y_n^m(\theta,\varphi)\Big| \leq \frac{n|q|}{\rho-\sqrt{3/4}a}\Big(\frac{\sqrt{3/4}a}{\rho}\Big)^{p+1} \equiv \epsilon_{me}.
\end{equation}

A key step in the FMM algorithm is the conversion of a ME about a given point into a local expansion about another point.  Suppose that $n$ ions of charge $q$ and spherical coordinates $(\rho_i,\theta_i,\varphi_i)$ are located in a cube of side length $d$ centered on a point $P_0=(\rho_0,\theta_0,\varphi_0)$ with $\rho_0>2\sqrt{3/4}a$.  Then the ME (centered on $P_0$) of this set of charges converges at any point $P=(\rho,\theta,\varphi)$ located in a cube of side length $a$ centered on the origin and the local Coulomb potential at $P$ is given by

\begin{equation}
\label{FMM_local}
\phi_c(\rho,\theta,\phi) = \sum_{n=0}^{\infty}\sum_{m=-n}^{n}L_n^mY_n^m(\theta,\varphi)\rho^n.
\end{equation}

This is equivalent to theorem 2.4 in \cite{cheng1999fast} except that the source and target regions are cubes. The local expansion (LE) coefficients $L_n^m$ depend only on the ME coefficients ($M_n^m$'s) and the point $P_0$.  Furthermore, the LE truncation error is similar to the ME error ($\epsilon_{le}\sim \epsilon_{me}$).  The efficient calculation of the LE coefficients from the ME coefficients is an involved process which is described in \ref{Plane-wave expansions}.

\subsubsection{Algorithm}

The FMM algorithm consists of calculating multipole and local expansions of an $N$ ion distribution at various levels of spatial refinement in order to accelerate the Coulomb force calculation. First, the smallest cube containing all $N$ ions is found.  Next, this cube is recursively divided into octants such that, after $k$ iterations, the space has been divided into a maximum of $8^k$ cubes.  This is a maximum value because the FMM3D library implements an adaptive routine in which a given cube is only further divided if it contains at least some minimum number of ions.  However, for the rest of this section, we will ignore details of the adaptive algorithm, for simplicity.  A more complete description of the FMM algorithm is found in \cite{greengard1997new,cheng1999fast}.

The remainder of the algorithm can be divided into two parts: the upward pass, in which MEs of the cubes are calculated for increasingly larger boxes and the downward pass, in which LEs are calculated for increasingly smaller boxes, allowing eventually for the calculation of the Coulomb force on each ion.  The upward pass begins at the finest spatial refinement level, which we will assume includes $8^f$ cubes.  The value of $f$ used depends on the configuration of ions and the required precision $\epsilon$.  The ME is calculated for each cube about its center (see equation \ref{FMM_coul}).  The exact number of terms calculated in the MEs depends on $\epsilon$. Next, groups of eight adjacent cubes are merged such that only $8^{f-1}$ cubes remain, and the MEs of these larger cubes are calculated. Importantly, the ME of a cube at this level is not found from scratch, but by using the MEs at the prior level. The MEs of the smaller cubes are translated to the center of the larger cube and the MEs are then summed.  This process is repeated recursively, generating the multipole expansions of increasingly larger cubes.  Details of the multipole translation operator, including its decomposition into rotations and translations,  can be found in the references.  

The next step is the downward pass.  Starting at the coarsest refinement levels, the LE about the center of each cube is calculated by converting the ME of distant cubes, according to equation \ref{FMM_local}.  Starting at refinement level 2 (with $8^2$ cubes), there exist cubes distant enough from one another to satisfy the hypothesis of equation \ref{FMM_local} and the LEs of each cube, due to distant cubes, are calculated.  Note that when forming the LE for a given cube, the potential due to its adjacent cubes is not included, since they are not far enough apart to satisfy the hypothesis.  Next, each "parent" cube is further divided into eight "child" cubes, forming refinement level 3 ($8^3$ total cubes).  The LE of a child cube, denoted $C$, is formed by adding two contributions.  First, the LE of the parent cube is translated to the center of $C$.  Second, there are now cubes which were not accounted for at level 2 which are now far enough away from $C$ to satisfy the hypothesis of equation \ref{FMM_local}.  The MEs of these distant cubes are converted into LEs about $C's$ center. A key point is that the cubes whose MEs are used here are distant enough from $C$ that the mathematical formalism can be used, but still relatively close to $C$.  Very distant regions have already been accounted for at the less refined level 2, which underscores the efficiency of the FMM algorithm.  At this point, the cubes are further divided and this process repeats through the finest level of refinement. Now, the LEs about the center of each cube at the finest level are known.  The potential at the location of each ion is then easily computed.  At this point, the only interactions unaccounted for are those between nearby ions, and these are calculated directly.  Note that this description has ignored certain details such as adaptive algorithms and that, when computationally faster, the direct calculation is sometimes used in place of the ME to LE conversion.  For large $N$, this algorithm has an operation count which scales directly with $N$.  The exact number of operations also depends on the number of terms used in the multipole expansions and the number of refinement levels, and is further discussed in the references.

\subsubsection{Plane-wave expansions}
\label{Plane-wave expansions}

It turns out, for three dimensional charge distributions, it is inefficient to convert multipole expansions directly to local ones or, in other words, to calculate the $L_n^m$'s in equation \ref{FMM_local} directly from the $M_n^m$'s in equation \ref{FMM_coul}.  A faster approach involves an intermediate conversion of the ME to a basis of plane-wave functions.  These plane-wave expansions (PWE) are then translated to the cube at which the LE is sought and converted to an LE.  While the direct ME to LE conversions has a computation time which scales with $p^3$, the translation of the PWE requires only $O(p^2)$ operations.  This speedup is, in fact, the reason why the FMM is faster than the direct Coulomb calculation for even moderate values of $N$. Importantly, the error of a given PWE is bounded using known relations, allowing the precision level to be maintained throughout the algorithm.  Details of the PWE conversions can be found in the previously cited references. 

\section{\label{sec:sim_benchmarking}Simulation Benchmarking}

\begin{figure}
    \centering
    \begin{subfigure}{0.5\textwidth}
        \subcaption{}
        \centering
        \includegraphics[scale=0.27]{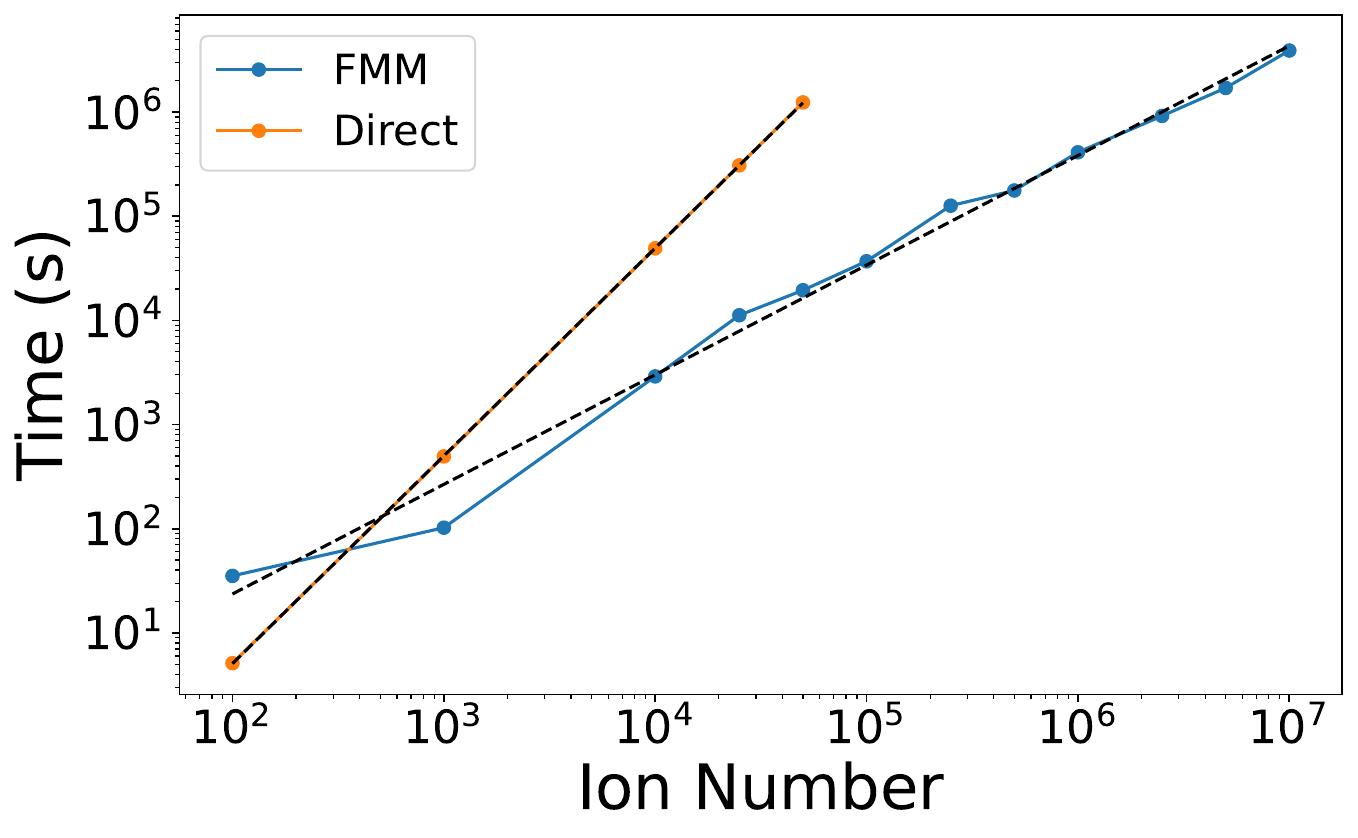}
        \label{fig:time_scaling}
    \end{subfigure}%
    ~ 
    \begin{subfigure}{0.5\textwidth}
        \subcaption{}
        \centering
        \includegraphics[scale=0.27]{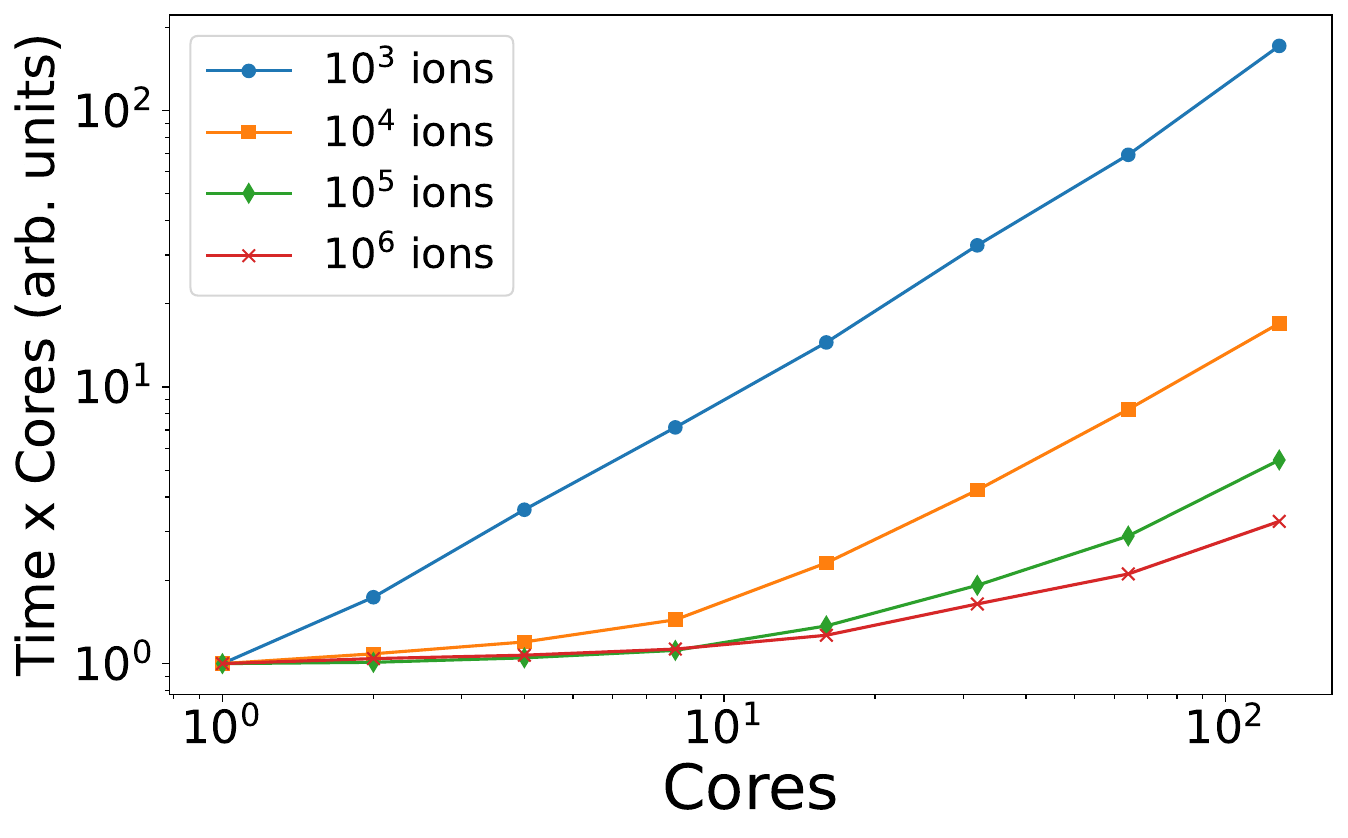}
        \label{fig:core_scaling}
    \end{subfigure}
    \caption{(a) Time required to simulate $10^4$ timesteps as a function of ion number, using either a direct Coulomb calculation or the FMM3D library.  The FMM method is faster than the direct one when simulating crystals of more than a few hundred ions.  The dotted black lines are fitted to the FMM and direct data and have slopes of 1.053 and 1.995, respectively, illustrating the $O(N)$ scaling of the FMM and $O(N^2)$ scaling of the direct calculation. (b) Strong scaling results for several different ion numbers.  The vertical axis uses units normalized such that the time corresponding to one core is equal to unity for each ion number considered.  Scaling improves significantly between $N=10^3$ and $N=10^4$.  For large ion number there exists nearly perfect strong scaling up to 8 cores.}
\end{figure}

\begin{figure}
    \centering
    \begin{subfigure}{0.5\textwidth}
        \subcaption{}
        \centering
        \includegraphics[scale=0.27]{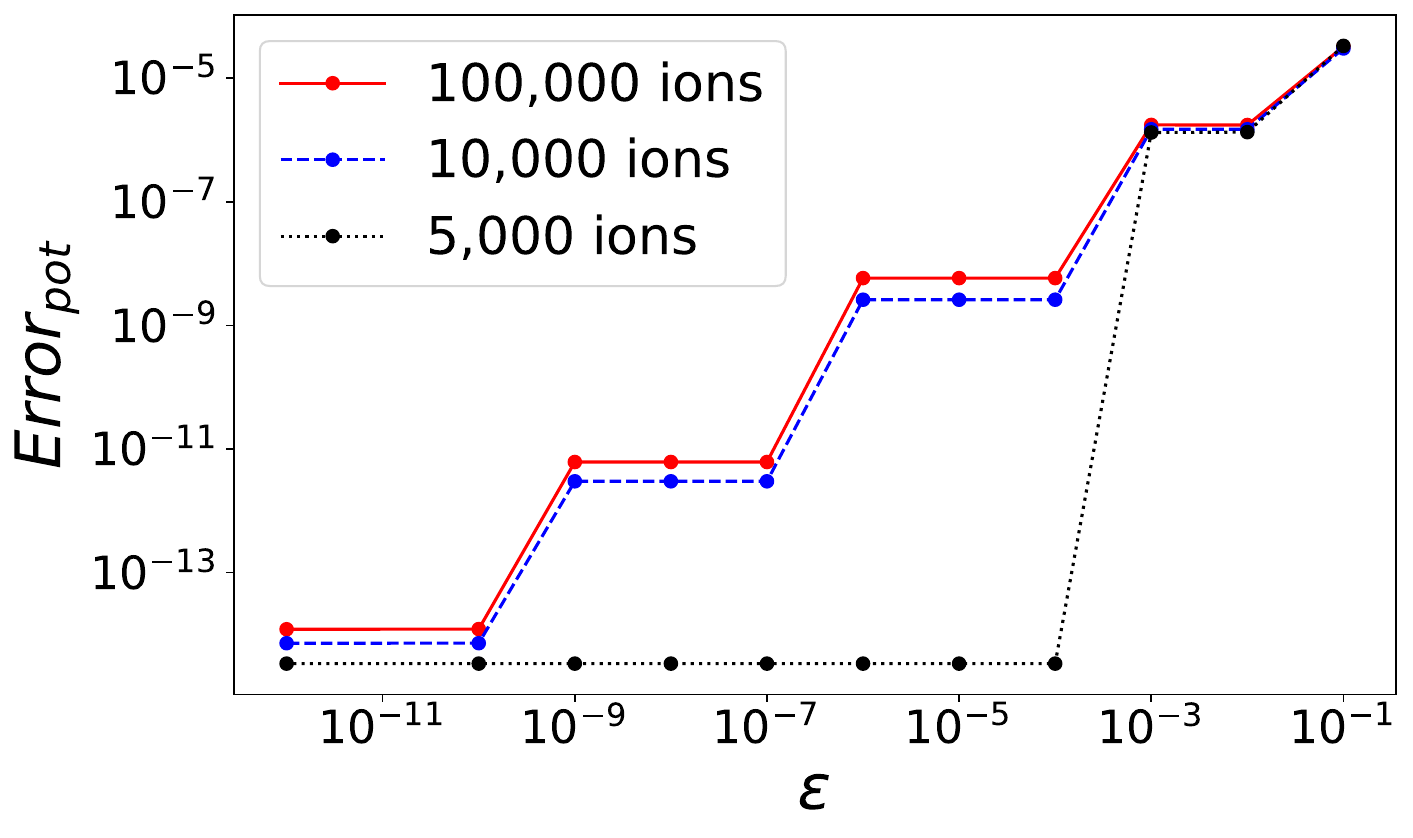}
        \label{fig:rmspot}
    \end{subfigure}%
    ~ 
    \begin{subfigure}{0.5\textwidth}
        \subcaption{}
        \centering
        \includegraphics[scale=0.27]{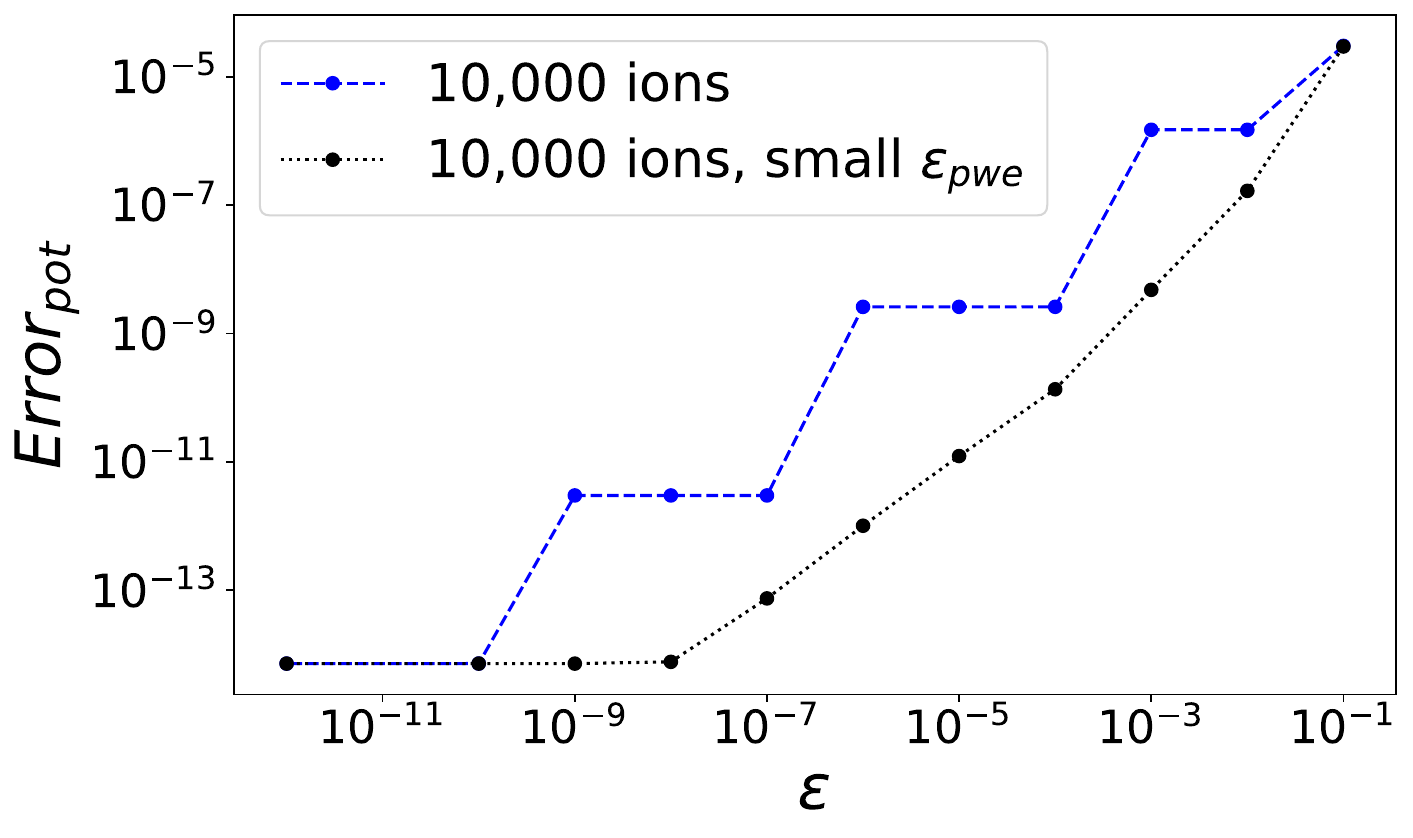}
        \label{fig:rmspot2}
    \end{subfigure}
    ~ 
    \bigskip
    \centering
    \begin{subfigure}{0.48\textwidth}
        \subcaption{}
        \centering
        \includegraphics[scale=0.27]{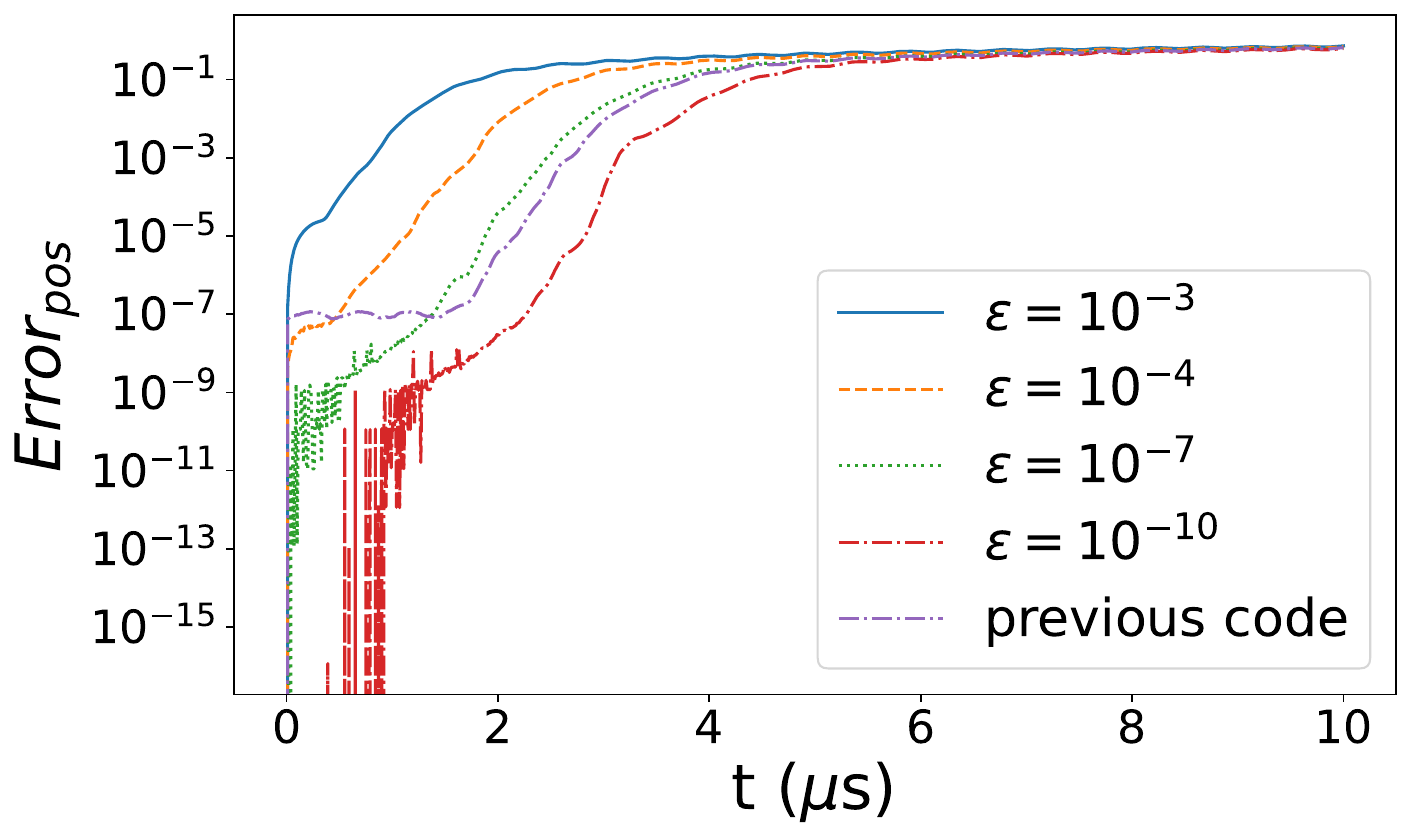}
        \label{fig:rmspos}%
    \end{subfigure}
    ~ 
    \begin{subfigure}{0.48\textwidth}
        \subcaption{}
        \centering
        \includegraphics[scale=0.27]{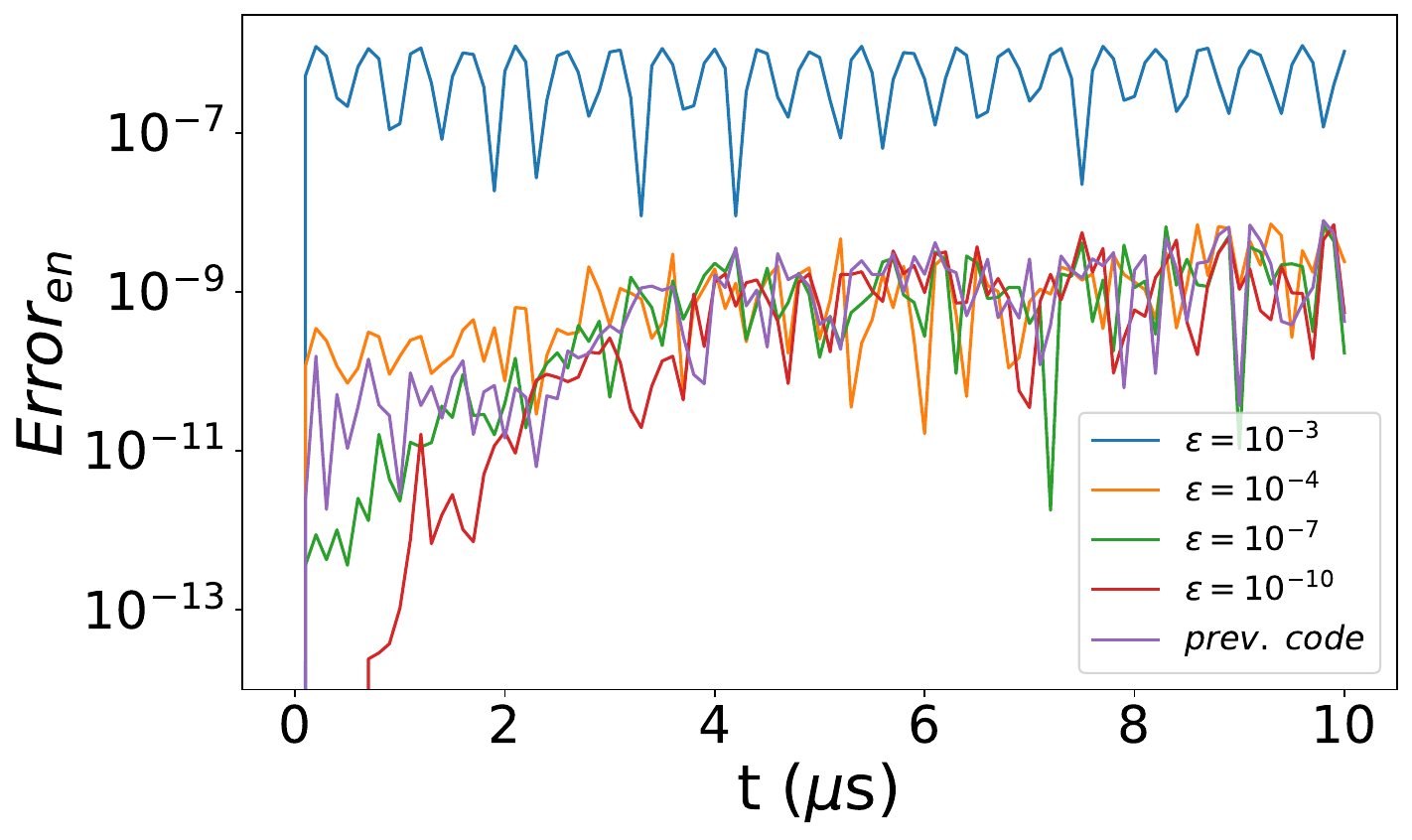}
        \label{fig:err_en}
    \end{subfigure}
    \caption{(a) Coulomb potential error using FMM3D library for various values of the precision parameter $\epsilon$.  The relative error of the Coulomb potential is computed by comparing the potential obtained using the FMM to that obtained from the direct calculation, according to equation \ref{rms_pot}. The FMM result converges to the direct one as $\epsilon$ decreases. As explained in the text, the precision of the FMM calculation, in fact, exceeds $\epsilon$. (b) Same as (a), except we compare the relative error with and without the changes to the FMM3D library discussed in the text, for $N=10^4$.  In particular, when the ME to PWE conversion is made more precise, the relative error decreases and becomes more sensitive to changes in $\epsilon$. (c) Relative error in ion positions over time, compared to the direct calculation, computed using equation \ref{rms_pos}.  In these simulations, $N=10^4$ ions are initialized with random positions within a sphere and allowed to evolve under the forces in the Penning trap.  Since the FMM approximates Coulomb forces, the difference in ion positions compared to the direct simulation grows over time.  The purple line shows the ion position error when comparing our new code to the previously published one \citep{tang2019first}.  The new simulation records ion positions to six decimal places which accounts for the purple line's plateau between 0 and 2 microseconds. (d)  Relative error in total energy over time, compared to the direct calculation, computed using equation \ref{rms_en}.  The energy error is much smaller than the ion position error and energy is effectively conserved even when $\epsilon$ is relatively large.  This validates the use of our code to study the time evolution of ion crystals. }
    \label{fig:convergence}
\end{figure}

We have performed several studies to demonstrate the improved efficiency of the code resulting from the use of the FMM3D library.  First, we observe linear scaling between simulation time and ion number.  Figure \ref{fig:time_scaling} shows the time required to simulate the evolution of a system of ions over $10^4$ timesteps as a function of ion number.  We did not initialize the ions in their cold crystal equilibria since it would be computationally difficult to calculate such equilibria for the larger ion numbers considered. Rather, for these tests, we initialized the ions by uniformly distributing them within a sphere such that the ion number density is the same for each ion number at $t=0$. The initial velocities are set to zero and the ion system evolves according to the Hamiltonian given in equation \ref{hamiltonian}. Simulation times using the direct Coulomb force calculation and the FMM3D library are compared.  The direct Coulomb simulation time scales with $N^2$, while the FMM simulation time scales with $N$ when $N$ is sufficiently large.

Since the FMM3D library implements shared-memory parallelism, we also studied strong scaling with the number of processor cores.  This is demonstrated in figure \ref{fig:core_scaling}, where we plot the product of the simulation time and number of cores versus the number of cores for several ion numbers.  We observe that, as the number of ions increases,  the scaling improves. For instance, when $N\gtrsim 10^5$ ions, the simulation time using 8 cores is $\sim 8$ times smaller than when using 1 core.  This is known as linear scaling and represents the ideal speedup.  For $>8$ cores, the scaling is no longer linear but the simulation time continues to decrease with the number of cores.

As explained in section \ref{sec: fmm}, the FMM simulation time also depends on the required precision of the electrostatic potential, denoted $\epsilon$, which in this section refers to the parameter passed to the FMM3D library.  It is important to choose a value of $\epsilon$ which precisely reproduces the direct calculation, but an unnecessarily small value can slow the FMM calculations.  Figure \ref{fig:convergence} demonstrates the convergence of the FMM simulation results to the direct Coulomb results as $\epsilon$ decreases.  In figure \ref{fig:rmspot} we plot a measure of the relative error in the electrostatic potential for different values of $\epsilon$.  Ions are randomly placed within a sphere and the Coulomb potential is calculated at the location of each ion due to all other ions.  Letting $\phi^{fmm}_i$ and $\phi^{dir}_i$ represent the Coulomb potential at the location of ion $i$ as found by the FMM and direct calculations, respectively, we define the relative error as

\begin{equation}
\label{rms_pot}
Error_{pot} = \sqrt{\frac{\sum_{i=1}^N|\phi^{fmm}_i-\phi^{dir}_i|^2}{\sum_{i=1}^N|\phi^{dir}_i|^2}}.
\end{equation}

In figure \ref{fig:rmspot}, the relative error is always significantly smaller than $\epsilon$. This is a result of using a conservatively large number of terms $p$ in the multipole expansions (i.e. \ref{FMM_coul}), which leads to $\epsilon_{me} < \epsilon$. Furthermore, the error exhibits a step-function-like behavior in which the relative error is constant over various ranges of $\epsilon$.  This is also a result of the precision of the various expansions used in the FMM, but is not explained by our previous discussion in section \ref{sec: fmm}.  Specifically, in comparison to the MEs, which are much more precise than needed, the ME to PWE conversions are not as precise, and therefore limit the overall error.  Furthermore, there exist various parameters in the PWE computation which are tabulated only for discrete values of $\epsilon$ \citep{cheng1999fast}, which causes the error to remain nearly constant over different ranges of $\epsilon$.  It is possible to edit the FMM3D library to obtain a more sensitive scaling between $Error_{pot}$ and $\epsilon$.  This is done by ensuring that the ME to PWE conversion error is smaller than the ME error, $\epsilon_{me}$. In figure 3b, we compare the relative error in the electric potential with and without this change to the code.  We see that, with the change, the step-function-like behavior vanishes and the error decreases with each decrease in $\epsilon$, until the ultimate precision is reached. In both cases the obtained error is less than $\epsilon$.

Using the FMM to calculate Coulomb interactions leads to errors in the forces exerted on the ions. This means that when running FMM and direct Coulomb simulations with identical initial conditions, the corresponding ion coordinates deviate over time.  However, while the exact microstate of the crystal will not be the same, bulk properties which characterize the macrostate of the crystal, such as its total energy, will agree with high precision.  These features are demonstrated in figure 3c, where we plot the error in ion positions versus time, and in figure 3d, where we plot the error in total energy versus time. In these simulations, $N=10^4$ ions are randomly placed within a sphere and allowed to evolve under the Hamiltonian in equation \ref{hamiltonian}.  Defining the ion position vectors in the FMM and direct simulations as $\boldsymbol{r}^{fmm}$ and  $\boldsymbol{r}^{dir}$, respectively, and the total ion crystal energies as $E^{fmm}$ and $E^{dir}$, we define the relative errors using

\begin{equation}
\label{rms_pos}
Error_{pos} = \sqrt{\frac{\sum_{i=1}^N|\boldsymbol{r}^{fmm}_i-\boldsymbol{r}^{dir}_i|^2}{\sum_{i=1}^N|\boldsymbol{r}^{dir}_i|^2}},
\end{equation}

\begin{equation}
\label{rms_en}
Error_{en} = \frac{|E^{fmm}_i-E^{dir}_i|}{|E^{dir}_i|}.
\end{equation}

Note that the energies are calculated in the rotating frame, or after applying the coordinate transformation in equation \ref{rotation}.    As expected, the position errors take longer to accumulate when $\epsilon$ is small, but still approach $O(1)$ in all cases, showing that the crystals simulated with the direct and FMM techniques have different ion positions after a few microseconds.  The error in total energy, on the other hand, is very small, even for relatively large values of $\epsilon$, suggesting that the FMM can be used to precisely study the overall properties of the crystal. In figures \ref{fig:rmspos} and \ref{fig:err_en}, in addition to comparing the FMM simulation to the direct one, we also compare the direct code to our group's previously published Penning trap simulation code \citep{tang2019first}.  These two codes are conceptually identical except that, as previously mentioned, the code used in this paper can switch between the direct and FMM-accelerated modes.  The main difference between the codes is that the one used in this paper is compiled and written entirely in C++, whereas the previous code uses Cython to connect various layers of code written in either Python or C++. The precise agreement between these codes helps to validate the rest of the simulations presented in this paper.  Note that, in most of the simulations presented in the remainder of this paper, we choose to set $\epsilon=10^{-7}$, which leads to an effective combination of computational precision and speed.

\section{3D Crystal Dynamics}
\label{sec:3D crystal dynamics}

\begin{figure}
    \centering
    \begin{subfigure}{0.5\textwidth}
        \subcaption{}
        \centering
        \includegraphics[scale=0.2]{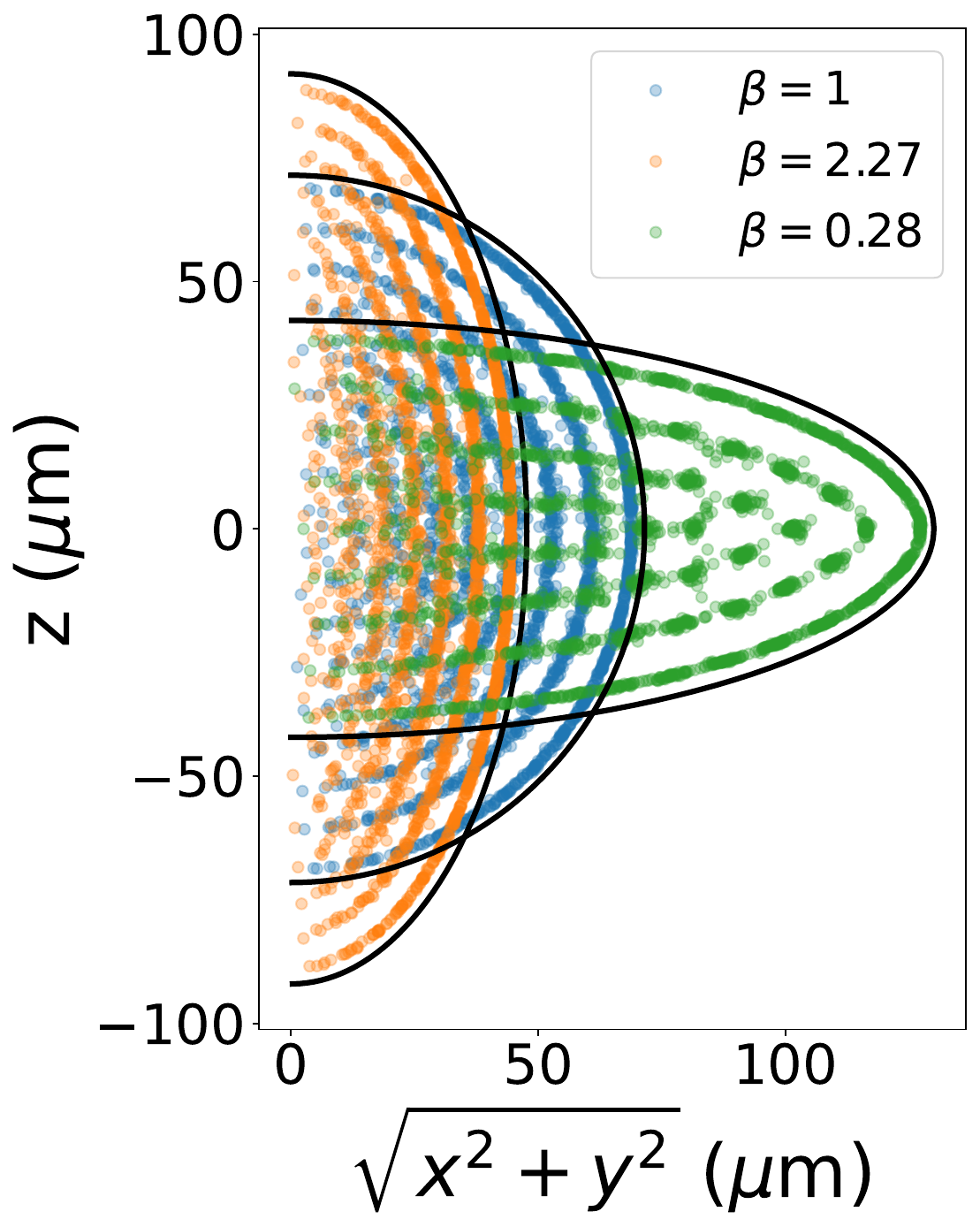}
        \label{fig:cry_shape}
    \end{subfigure}%
    ~ 
    \begin{subfigure}{0.5\textwidth}
        \subcaption{}
        \centering
        \includegraphics[scale=0.2]{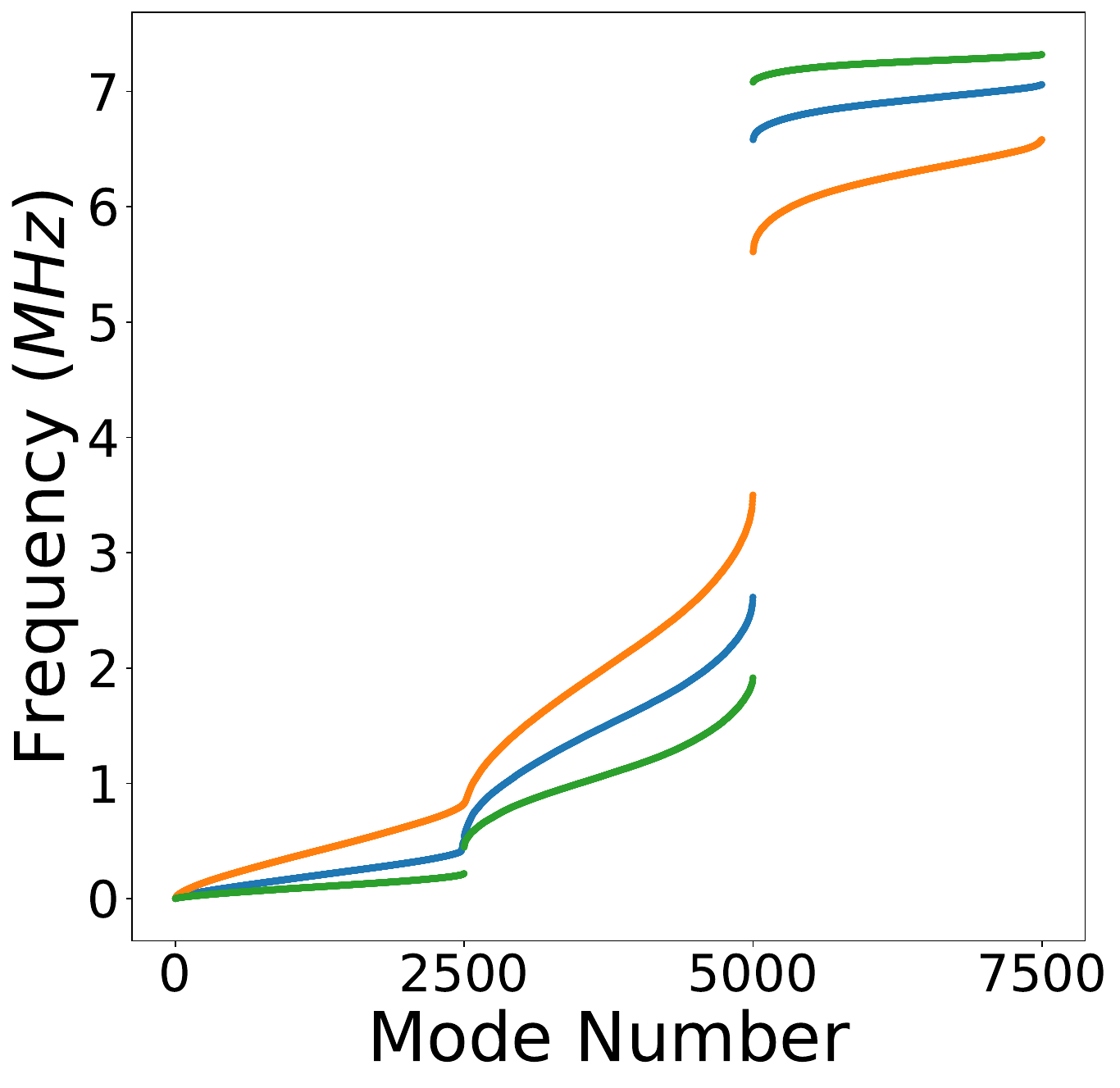}
        \label{fig:mode_spec}
    \end{subfigure}
    ~ 
    \bigskip
    \centering
    \begin{subfigure}{0.5\textwidth}
        \subcaption{}
        \centering
        \includegraphics[scale=0.2]{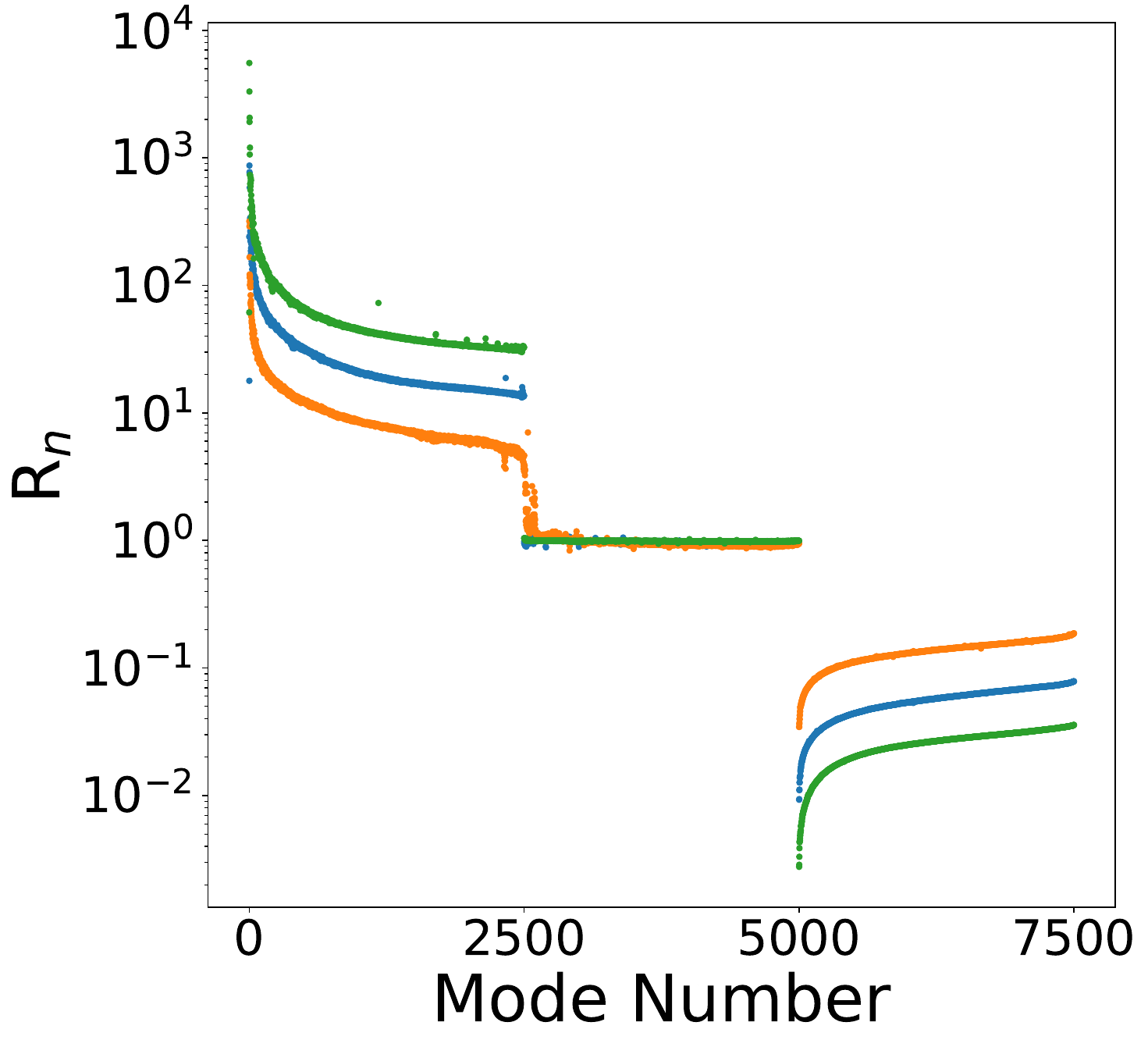}
        \label{fig:Rn}%
    \end{subfigure}
    ~ 
    \begin{subfigure}{0.45\textwidth}
        \subcaption{}
        \centering
        \includegraphics[scale=0.2]{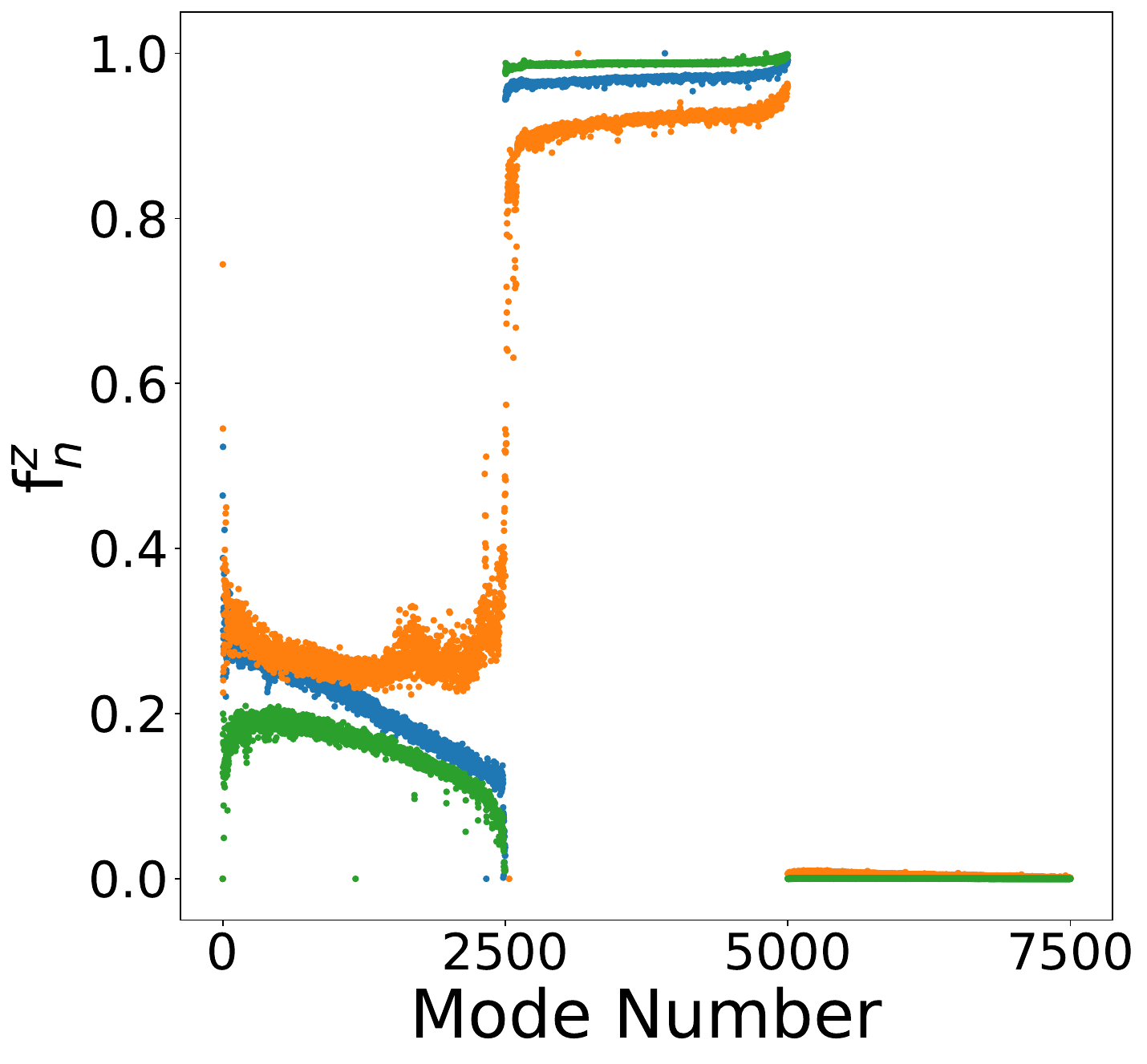}
        \label{fig:fn}
    \end{subfigure}
    \caption{(a) Equilibria of $N=2500$ ion crystals for various values of rotating wall frequency $\omega_r$, found using numerical energy minimization methods.  Plotted are the $z$ coordinates versus the cylindrical radius from the center of the trap for each ion.   As $\omega_r$ increases, so does the shape factor $\beta$, which elongates the crystal in the axial direction. Solid black lines correspond to the theoretical ion crystal shapes, found using equation \ref{eq:shape}.  (b)  Mode structure of the ion crystals shown in (a).  For small values of $\beta$, the mode frequencies are separated into three rather distinct branches, similar to what is seen in 2D crystals.  As $\beta$ increases, the gaps between the branches decrease.  The mode branches, in order of increasing frequency, correspond to the $\boldsymbol{E}\times\boldsymbol{B}$, axial, and cyclotron modes. (c) Ratio of potential to kinetic energy of each eigenmode, calculated using equation \ref{eq:pe_comp}.  For the cases considered here, the $\boldsymbol{E}\times\boldsymbol{B}$ (cyclotron) modes become more potential (kinetic) energy dominated as $\beta$ decreases.  Most axial modes are nearly harmonic for all three values of $\beta$. (d) Axial component of each eigenmode, calculated using equation \ref{eq:z_comp}. As $\beta$ increases, the $\boldsymbol{E}\times\boldsymbol{B}$ modes, which are no longer purely planar, become more axial in character, while the 'axial' modes gain a planar component.}
\end{figure}

The motion of ion crystals near their equilibrium configurations is an important topic of research with applications in a wide range of experiments.  Therefore, before studying the dynamics of a 3D ion crystal, it is necessary to first calculate its equilibria by minimizing the rotating frame potential energy.  Our simple energy minimization technique consists of first finding an ion configuration corresponding to a local minimum of the potential energy, then nudging the ion positions and repeating the local minimization.  This process is repeated several times and the lowest energy configuration found is returned.  We use the SciPy library's BFGS minimization to calculate the local minimum ion configurations. It should be noted that each computation of the potential and its gradient which occurs during the minimization makes use of the FMM3D library in order to accelerate the calculation.  Unfortunately, efficient methods to find the global energy minimum of 3D crystals are not available, and it is not guaranteed that we find such a configuration using our minimization procedure.  In figure \ref{fig:cry_shape}, we plot the equilibrium ion positions for  an $N=2500$ ion crystal using various trap parameters. Since $\beta$ depends on $\omega_r$ (equation \ref{beta}), we are able to achieve crystals with different shapes by varying $\omega_r$. 

While it is computationally challenging to find the minimum energy ion configuration, the gross shape of the crystal is well-understood theoretically.  While $\beta$ describes the relative axial and planar crystal dimensions, there also exists cylindrical asymmetry, parameterized by $\delta$ in equation \ref{pot_wall}, due to the rotating wall. This leads to the ion crystal being stretched and compressed along perpendicular directions in the plane.  The shape of the resulting ellipsoid ion crystal (assumed to have uniform density) is fully described by the semi-axes $(a_x,a_y,a_z)$.  We define $\{a_1,a_2,a_3\}=\{a_x,a_y,a_z\}$ with $a_1>a_2>a_3$, noting that the mapping from $\{a_x,a_y,a_z\}$ to $\{a_1,a_2,a_3\}$  is based on the order of the coefficients $\{C_x,C_y,C_z\}$ in equation \ref{eq:pe_rot}.  As an example, if $C_x>C_y>C_z$, then $a_1=C_z$, $a_2=C_y$, and $a_3=C_x$.   One must then solve two of the following equations for the ratios $a_2/a_1$ and $a_3/a_1$ \citep{binney2011galactic, dubin1999trapped}:

\begin{subequations}
\label{eq:shape}
\begin{align}
\omega_r(\omega_c-\omega_r)&\frac{a_2a_3}{a_1^2}\frac{F(\theta, k)-E(\theta,k)}{k^2\sin^3\theta} = \omega_z^2\cdot\min\{C_x,C_y,C_z\}\label{shape1}\\  
\omega_r(\omega_c-\omega_r)&\frac{a_2a_3}{a_1^2}\frac{E(\theta, k)-(1-k^2)F(\theta,k)-(a_3/a_2)k^2\sin\theta}{k^2(1-k^2)\sin^3\theta} = \omega_z^2\cdot \textrm{median}\{C_x,C_y,C_z\}\label{shape2}\\  
\omega_r(\omega_c-\omega_r)&\frac{a_2a_3}{a_1^2}\frac{(a_2/a_3)\sin\theta-E(\theta,k)}{(1-k^2)\sin^3\theta} = \omega_z^2\cdot\max\{C_x,C_y,C_z\}.\label{shape3}
\end{align}
\end{subequations}

Here, $k^2 = (a_1^2-a_2^2)/(a_1^2-a_3^2)$, $\theta = \cos^{-1}(a_3/a_1)$, and $F$ and $E$ denote elliptical integrals of the first and second kind, respectively. For the crystals in figure \ref{fig:cry_shape} we choose $\delta<<\beta$ so that there is little anisotropy in the plane.  This leads to a clear illustration of concentric shells of ions and allows the theoretical shape of each crystal (from equation \ref{eq:shape}) to be illustrated with a single curve.  We include the predicted shapes of the crystals as black curves in figure \ref{fig:cry_shape} and they agree well with the calculated crystal equilibria.  Note that equation \ref{eq:shape} only provides the ratios of the ellipsoid axes, so we fit the absolute size of the predicted ellipsoid to match that of the crystal.

The normal modes of an ion crystal are obtained by first expanding the system's Lagrangian about its equilibrium configuration and deriving the Euler-Lagrange equations of motion \citep{wang2013phonon}.  Details of the calculation are provided in Appendix \ref{appA} and similar treatments have been applied in recent studies of planar crystals \citep{jain2020scalable,shankar2020broadening}, bilayer crystals \citep{Hawaldar:2023czr}, and 3D crystals \citep{dubin2020normal}.

Introducing the state vector $\boldsymbol{q} = (\boldsymbol{\delta r}, \boldsymbol{v})^T$, which contains the displacements from equilibrium and velocities of each ion, the Euler-Lagrange equations of motion near equilibrium (in the rotating frame) are found to be

\begin{equation}
\label{shape}
\frac{d\boldsymbol{q}}{dt} = \begin{pmatrix}
    \mathsfbi{0}_{3N}&\mathsfbi{1}_{3N}\\-\mathsfbi{K}/m&-2\mathsfbi{W}/m 
    \end{pmatrix}\boldsymbol{q}\equiv \mathsfbi{D}\boldsymbol{q}.
\end{equation}

In the above matrix, each element itself is a $3N\times3N$ square matrix. $\mathsfbi{K}$ is known as the stiffness matrix and is real and symmetric while $\mathsfbi{W}$ accounts for the Lorentz force in the rotating frame and is a real, antisymmetric matrix. The matrix elements of $\mathsfbi{K}$ and $\mathsfbi{W}$ are given in Appendix \ref{appA}.  An eigenvector of the system has harmonic time dependence, so the $n^{th}$ eigenvector is $\boldsymbol{u_n}=\boldsymbol{u_n^0}e^{-i\omega_nt}$ and satisfies $\mathsfbi{D}\boldsymbol{u_n} = -i\omega_n\boldsymbol{u_n}$.  We numerically diagonalize $\mathsfbi{D}$ to calculate a crystal's eigenmodes.  While, in this formalism, state vectors like $\boldsymbol{u_n}$ are $6N$ dimensional, there are always only $3N$ linearly independent eigenvectors and positive eigenvalues.  In figure \ref{fig:mode_spec}, we illustrate how the mode spectrum of a 3D ion crystal changes as $\omega_r$ and, therefore, its shape, varies.  In each case, there appear to be three distinct mode branches which span different frequency ranges, although the frequencies of each branch and frequency spacing between branches change with $\omega_r$.  It has been shown that the mode spectrum of planar crystals similarly includes three branches    \citep{wang2013phonon} which, in order of increasing frequency, are known as the $\boldsymbol{E}\times\boldsymbol{B}$ (or magnetron), axial, and cyclotron modes. Following \cite{Hawaldar:2023czr} we will refer to the 3D crystal mode branches using the same names as their planar crystal analogues, even though they exhibit certain qualitative differences.

The various eigenmodes of a Penning trap ion crystal do not all behave as simple harmonic oscillators.  In particular, due to the presence of a magnetic field, a given eigenmode's time-averaged kinetic energy is not necessarily equal to its time-averaged potential energy. The ratio of the time-averaged potential and kinetic energy of modes has been studied in 2D \citep{shankar2020broadening} and bilayer 3D \citep{Hawaldar:2023czr} crystals.  In figure \ref{fig:Rn}, we plot this ratio for our ellipsoid crystals, calculated using

\begin{equation}
\label{eq:pe_comp}
R_n = \frac{\boldsymbol{u_n^{r\dagger}}\mathsfbi{K}\boldsymbol{u_n^r}}{m\boldsymbol{u_n^{v\dagger}}\boldsymbol{u_n^v}}.
\end{equation}

Here, $\boldsymbol{u_n^r}$ and $\boldsymbol{u_n^v}$ are the $3N$-dimensional position and velocity components of the $n^{th}$ eigenmode, respectively, and the dagger symbol represents the conjugate transpose.  For each 3D crystal shown, most axial modes contain nearly equal kinetic and potential energy contributions. Similar behavior is seen in planar crystals, where the axial modes describe
simple harmonic motion and, therefore, contain exactly equal kinetic and potential energy contributions. On the other hand, the $\boldsymbol{E}\times\boldsymbol{B}$ (cyclotron) modes are potential (kinetic) energy dominated.  This is also true in the case of planar crystals. For the trap parameters used here, the $\boldsymbol{E}\times\boldsymbol{B}$ and axial mode energies become more equally divided between kinetic and potential as $\beta$ (or $\omega_r$) increases.  Qualitatively, this agrees with the results of \cite{Hawaldar:2023czr}, who compared the modes of a bilayer and planar crystal. It was found that the $R$ values of the $\boldsymbol{E}\times\boldsymbol{B}$ and cyclotron modes of the bilayer crystal, which had a larger value of $\beta$ than the planar crystal, were closer to unity.

Modes can also be characterized by their amplitudes in each spatial direction.  In planar crystals, for example, the $\boldsymbol{E}\times\boldsymbol{B}$ and cyclotron modes correspond to motion in the $xy$ plane while the axial modes describe strictly the $z$ motion of the ions.  On the other hand, the Coulomb interaction leads to mixing between the spatial degrees of freedom in 3D crystals, causing most modes to acquire both planar and axial components. In figure \ref{fig:fn} we plot the fraction of each mode which is in the axial ($z$) direction, given by

\begin{equation}
\label{eq:z_comp}
f^z_n = \frac{\boldsymbol{u_n^{z\dagger}}\boldsymbol{u_n^z}}{\boldsymbol{u_n^{r\dagger}\boldsymbol{u_n^r}}}.
\end{equation}

Here, $\boldsymbol{u_n^z}=(u_n^{z_1},.\;.\;.,u_n^{z_N})$ is the $N$ dimensional vector consisting of the $z$ components of the eigenvector corresponding to mode $n$.  While the cyclotron modes are still almost entirely planar, the $\boldsymbol{E}\times\boldsymbol{B}$ modes gain large axial components and the axial modes gain smaller, but still significant, planar components.  As may be expected, this mixing increases as the crystal's axial extent increases.  The crystal corresponding to $\beta>1$ exhibits axial modes with the greatest planar amplitudes and $\boldsymbol{E}\times\boldsymbol{B}$ modes with the greatest axial amplitudes. The behavior of the $\boldsymbol{E}\times\boldsymbol{B}$ modes is especially interesting.  For the $\beta=1$ crystal, and less so for the $\beta<1$ crystal, the highest frequency $\boldsymbol{E}\times\boldsymbol{B}$ modes have smaller axial components than the lowest frequency ones.  However, for the $\beta > 1$ crystal, the highest and lowest frequency $\boldsymbol{E}\times\boldsymbol{B}$ modes have comparably large axial components and it is actually the intermediate frequency modes which have slightly less axial character.

\section{\label{sec:laser_cooling}Laser Cooling}

\begin{figure}
    \centering
    \begin{subfigure}{0.5\textwidth}
        \subcaption{}
        \centering
        \includegraphics[scale=0.23]{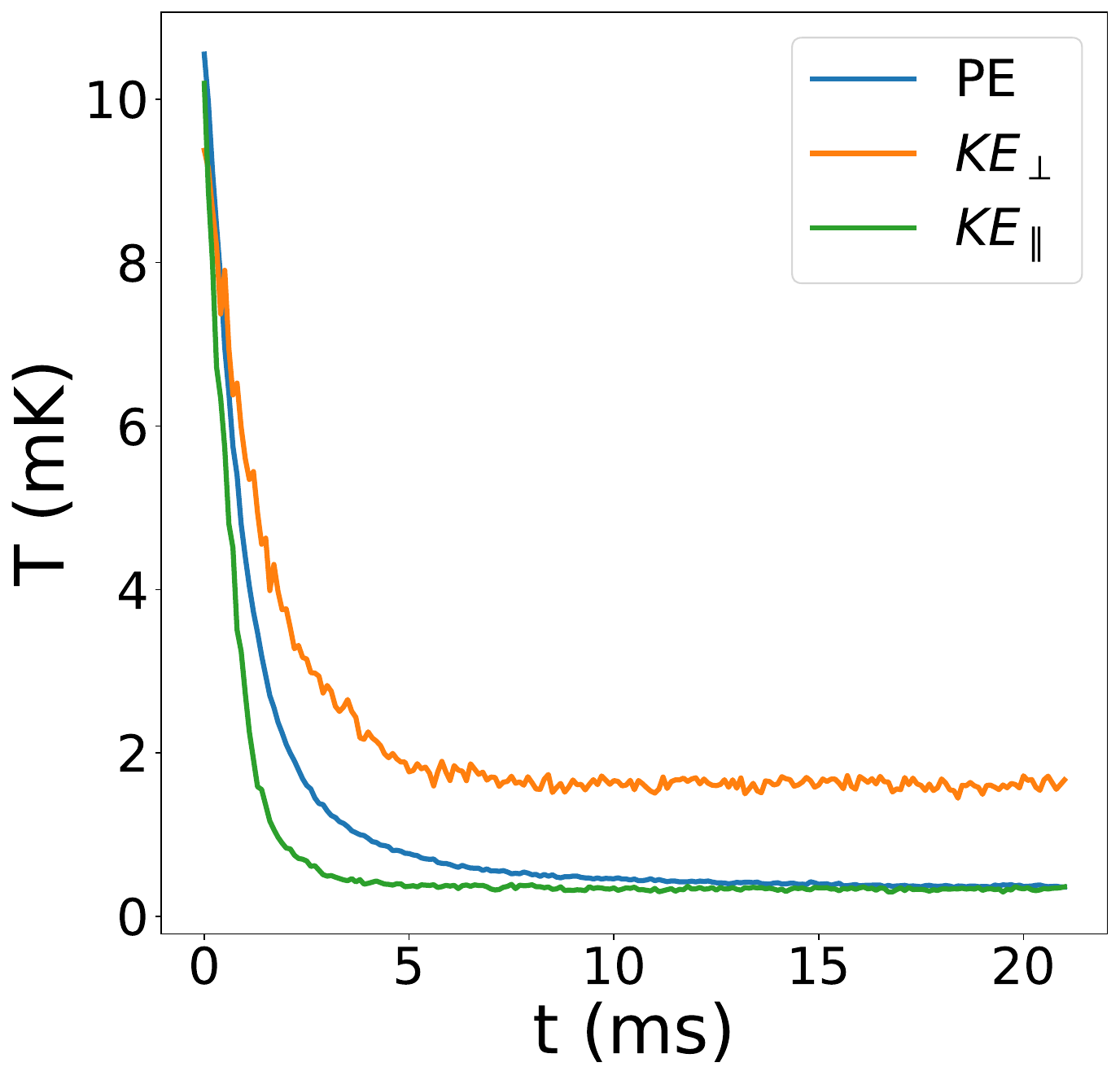}
        \label{fig:cooling_example}
    \end{subfigure}%
    \begin{subfigure}{0.5\textwidth}
        \subcaption{}
        \centering
        \includegraphics[scale=0.23]{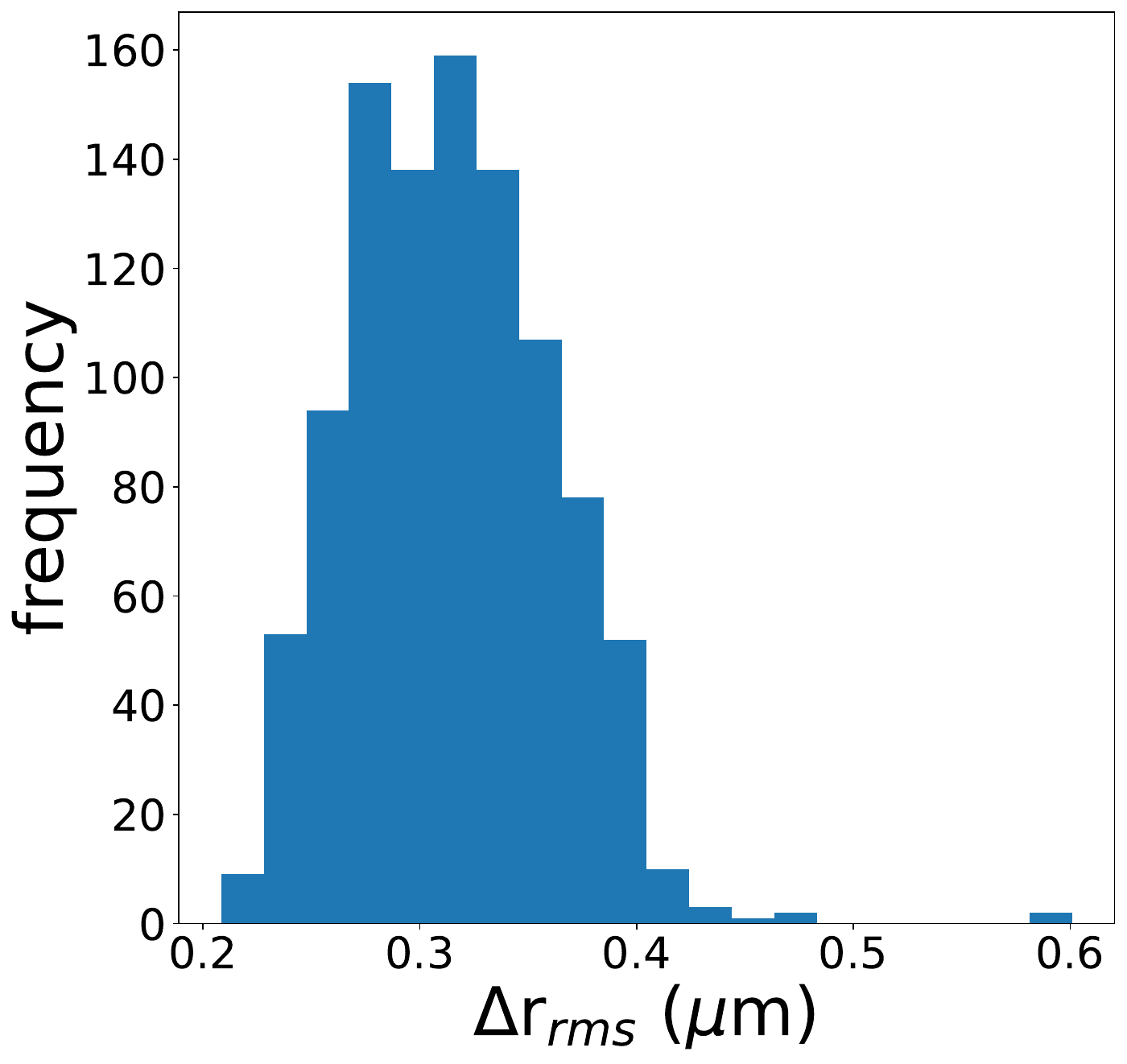}
        \label{fig:ion_disps}
    \end{subfigure}
    \caption{(a) Simulated cooling of kinetic and potential energies of the $N=1000$ ion crystal over 20 ms, while the planar and axial beams are turned on.  All temperatures are initialized at 10 mK, as described in the text.  In this simulation $w_y = 2.48\;\mu$m and $\Delta_{\perp} = 2.8$ MHz, while other laser parameters are provided in the text. (b) Histogram of the root mean square ion position fluctuations within 1 ms after the cooling lasers are turned off, for the same laser parameters as in (a). }
    \label{fig:cooling_single_sim}
\end{figure}

\begin{figure}
    \centering
    \begin{subfigure}{0.5\textwidth}
        \subcaption{}
        \centering
        \includegraphics[scale=0.23]{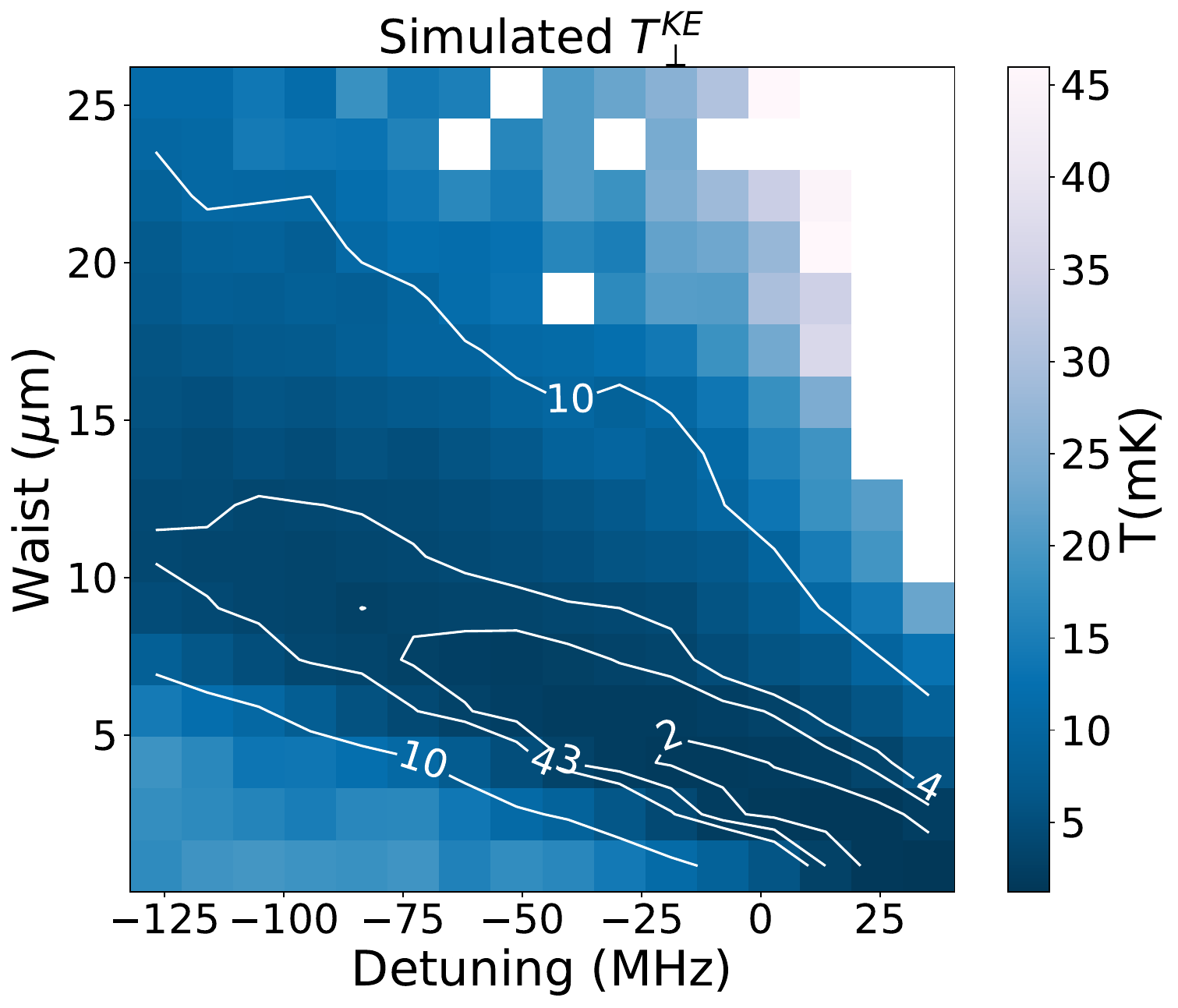}
        \label{fig:cooling_results_a}
    \end{subfigure}%
    ~ 
    \begin{subfigure}{0.5\textwidth}
        \subcaption{}
        \centering
        \includegraphics[scale=0.23]{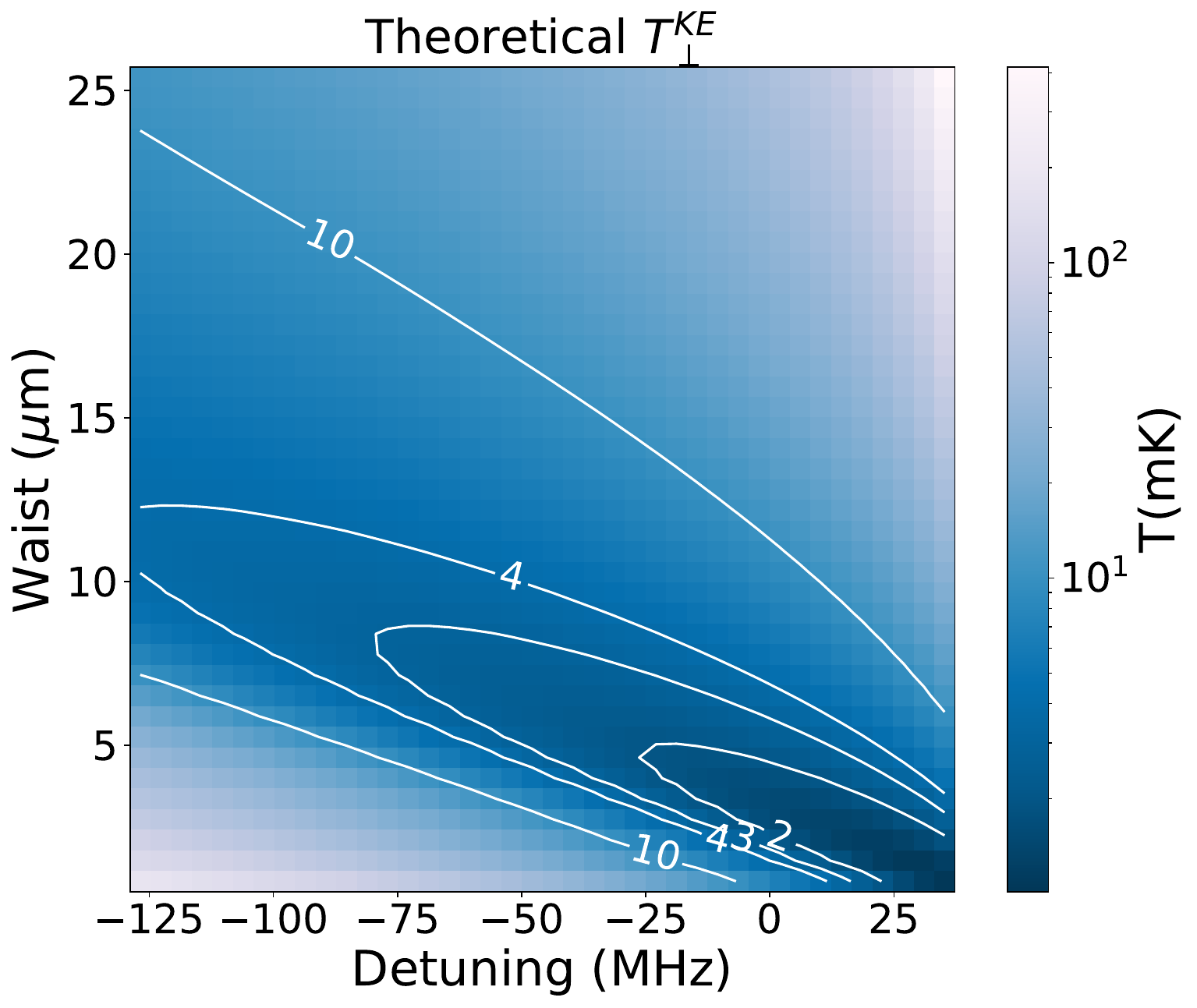}
        \label{fig:cooling_results_b}
    \end{subfigure}
    ~ 
    \bigskip
    \centering
    \begin{subfigure}{0.5\textwidth}
        \subcaption{}
        \centering
        \includegraphics[scale=0.23]{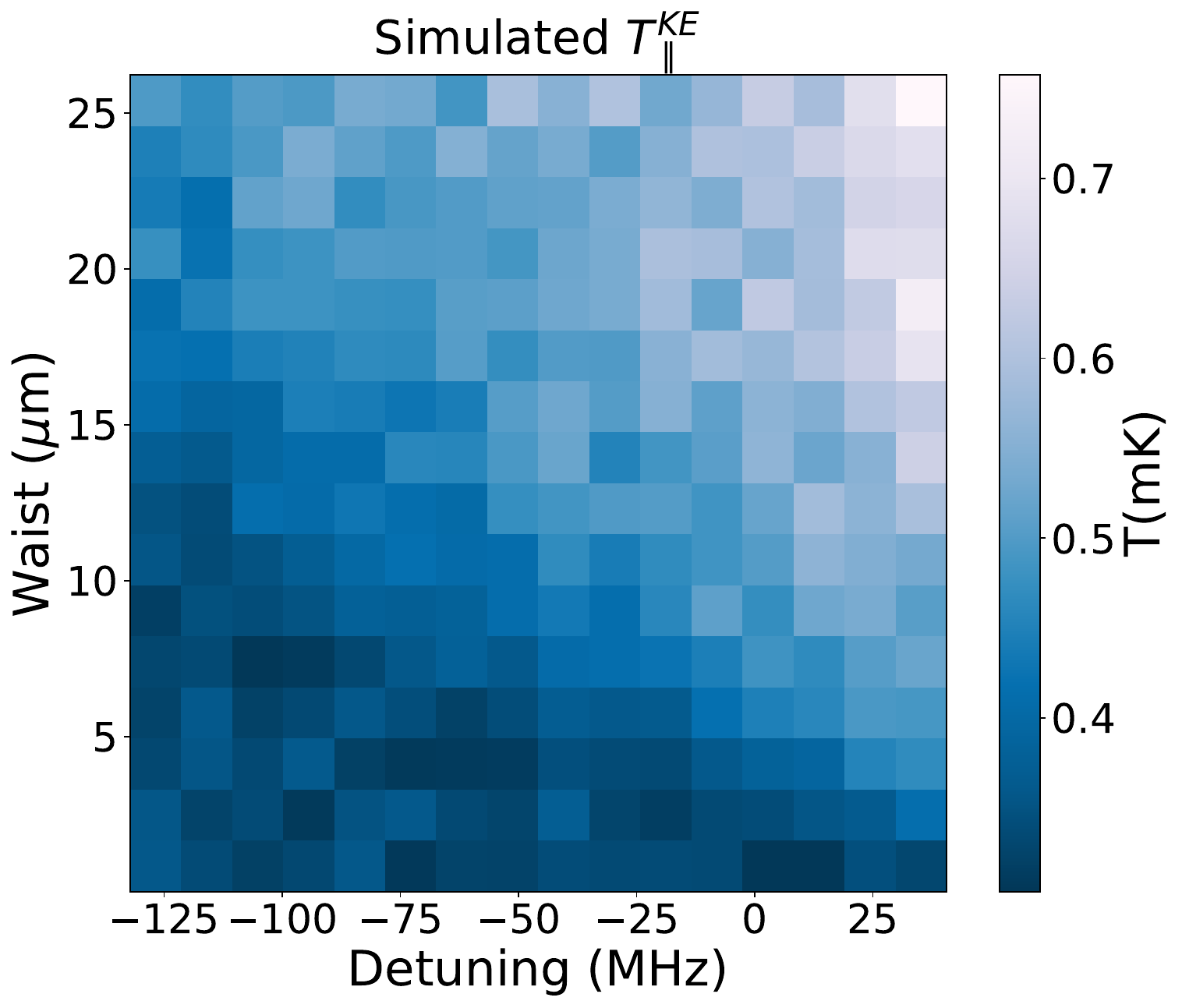}
        \label{fig:cooling_results_c}
    \end{subfigure}%
    ~
    \begin{subfigure}{0.5\textwidth}
        \subcaption{}
        \centering
        \includegraphics[scale=0.23]{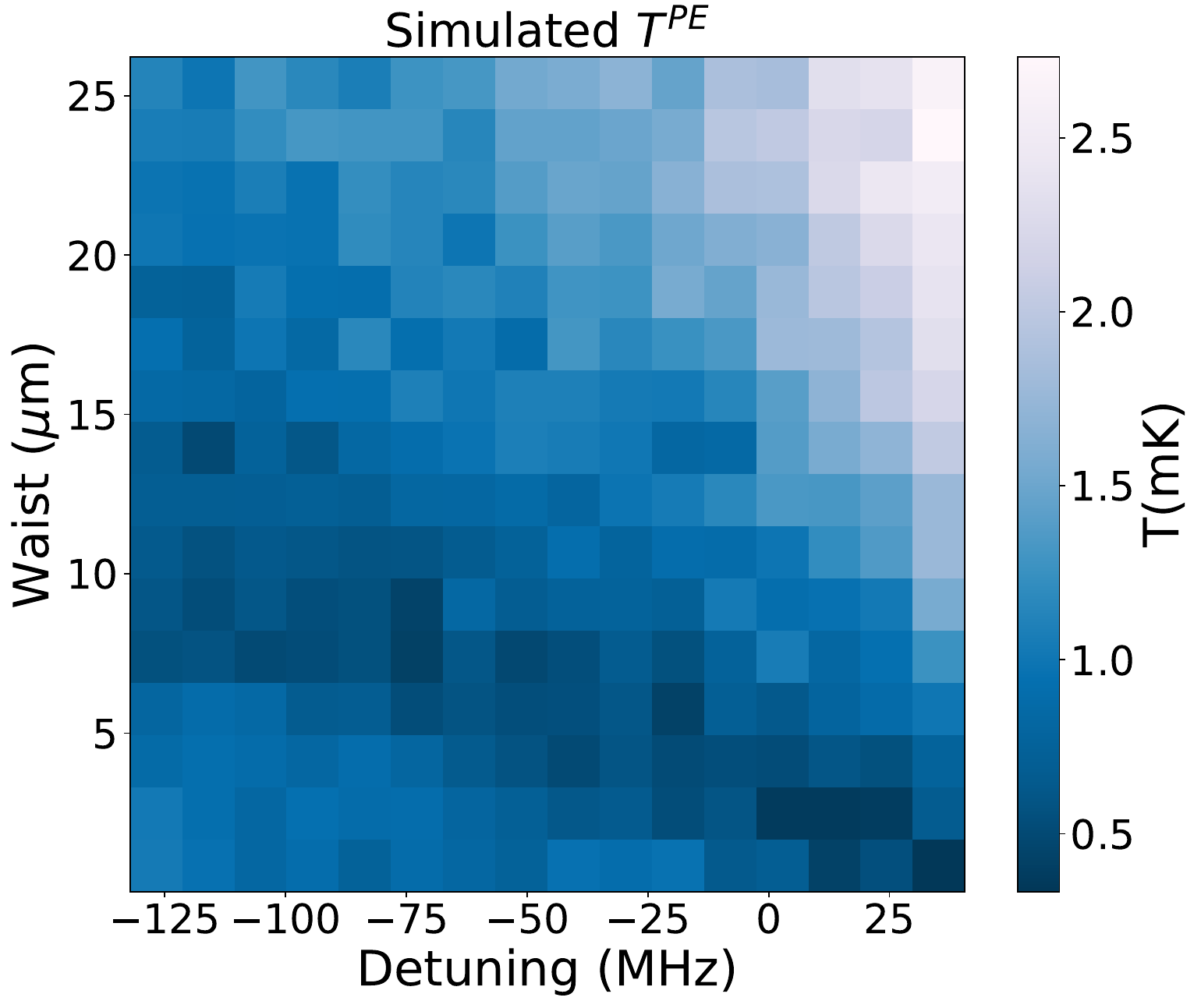}
        \label{fig:cooling_results_d}
    \end{subfigure} 
    \caption{(a) Simulated temperatures corresponding to the planar kinetic energy of the spherical $N=1000$ ion crystal, after 20 ms of laser cooling.  The simulations are identical to the one described in figure 5, except for the values of the beam waist $w_y$ and detuning $\Delta_{\perp}$ used in each run.  The blank boxes represent parameters for which the planar temperature did not equilibrate after 20 ms. (b)  Theoretical post-cooling temperatures corresponding to  the planar kinetic energy. (c) Simulated temperatures corresponding to the axial kinetic energy for the same set of parameters shown in (a) and (b).  Note that the post-cooling axial temperatures are below 1 mK for the entire parameter space shown. (d) Simulated temperatures corresponding to the total potential energy of the crystal. }
    \label{fig:cooling_results}
\end{figure}

A numerical simulation framework incorporating the FMM is exciting since it allows for the study of problems which were previously too large to simulate.  One of these problems is the laser cooling of large crystals. Here, we study the laser cooling of 3D ion crystals with $N=1000$ ions confined in a Penning trap.  From figure \ref{fig:time_scaling}, one can see that the FMM is faster than the direct method at $N=1000$ by a factor of $\sim 5$.  Future work will aim to study even larger crystals, for which the direct method is even less feasible.

The laser cooling of planar ion crystals in Penning traps has recently been a topic of interest, as obtaining ultracold temperatures is necessary for high fidelity quantum information and sensing protocols.  In the future, 3D crystals are expected to gain popularity as a tool for similar studies since the sensitivity of many quantum sensing experiments increases when more ions are interrogated. The laser cooling setup considered for 3D crystals is shown in figure \ref{cooling_setup} and described in section \ref{sec:theory}. The minimum achievable crystal temperature depends sensitively on the detuning, $\Delta_{\perp}$, of the planar beam from the cooling transition and, additionally, on the beam width in the $y$ direction, $w_y$. For 3D crystals, the width of this beam in the axial direction, $w_z$, must be considered, as well.  For instance, if $w_z$ is much smaller than the axial extent of the crystal, then cooling in the planar directions will likely be slow since most ions will not absorb photons from the planar beam.   For the purposes of this paper, we set $w_z$ to be large compared to the axial extent of the crystal to maximize the laser cooling efficiency.  We study the dependence of the ion crystal's post-cooling temperature on $w_y$ and $\Delta_{\perp}$, with a fixed beam offset $d = 5\;\mu$m and saturation parameter, or maximum saturation intensity, $S_{\perp,0}=0.5$.  The axial beams used in these simulations are characterized by $\Delta_{\parallel} = -\gamma_0/2$, $S_{\parallel,0} = 5\times 10^{-3}$, and uniform intensity across the extent of the crystal.

We first initialize a nearly spherical crystal of 1000 ions slightly out of equilibrium so that it can be laser cooled.  In particular, we set the initial kinetic and potential energies of the crystal to $T_i=10$ mK.  For a system of harmonic oscillators, it is possible to initialize all modes of the system at a temperature $T_i$ by simply initializing the kinetic energy at temperature $2T_i$ and the potential energy at temperature 0.  However, since the modes of an ion crystal in a Penning trap can have unequal average kinetic and potential energies, this cannot be done here \citep{shankar2020broadening}.  In particular, the $\boldsymbol{E}\times\boldsymbol{B}$ modes are potential energy dominated and, therefore, would have very small amplitudes if the ions were initialized with only kinetic energy.  Therefore, it is important to initialize both the kinetic and potential energy of the crystal.  The kinetic energy is initialized by randomly sampling the total speed, $v$, of each ion from the Maxwell-Boltzmann distribution, given by

\begin{equation}
\label{maxwell_boltzmann}
    f(v) = 4\pi\bigg(\frac{m}{2\pi k_bT_i}\bigg)^{3/2}v^2\exp\bigg(-\frac{mv^2}{2k_bT_i}\bigg).
\end{equation}
Here, $k_b$ is Boltzmann's constant.  In order to initialize the potential energy, we use a generalization of the Metropolis-Hastings algorithm described in \cite{shankar2020broadening}.  Starting from the ion crystal's equilibrium configuration $\{\boldsymbol{x}_{j}^0\}$, ions are sequentially displaced in order to increase the potential energy to a value typical of the set temperature $T_i$.  A given ion is randomly displaced in a random direction by a small distance $rand(0,\delta x)$, where we use $\delta x = 1\;\mu m$ for our 1000 ion crystal.  Denoting the potential energy of the crystal before and after this displacement as $E_{old}$ and $E_{new}$, respectively, the displacement is accepted if $E_{new}<E_{old}$. The change is also accepted if $rand(0,1) < \exp[-(E_{new}-E_{old})/k_bT_i]$.  Otherwise, the displacement is rejected and the ion is moved back to its previous position.   One scan of this algorithm is completed after this process is repeated for each ion and we complete 2000 scans in order to initialize the potential energy of the crystal.

After the crystal is initialized, its laser cooling is simulated.  Details of the laser cooling simulation methodology can be found in section \ref{sec:ion_dynamics}.  The decrease in kinetic and potential energies of a 3D ion crystal over time, for the laser parameters $(w_y, \Delta_{\perp}) = (2.48\;\mu$m, $13.6$ MHz$)$, is illustrated in figure \ref{fig:cooling_example}.  It can be seen that the kinetic energy in the axial direction and the potential energy ($KE_{\parallel}$ and $PE$, respectively) cool to below 1 mK within a few milliseconds whereas the kinetic energy in the plane is cooled, but remains at a significantly higher temperature.  The temperatures corresponding to the axial kinetic, planar kinetic, and potential energies are found using the following formulas:

\begin{subequations}
\label{eq:temps}
\begin{align}
T^{KE}_{\parallel} &= \frac{m\sum_{i=1}^N\langle\dot{z}_i^2\rangle}{Nk_b},\label{eq:temp_ax}\\  
T^{KE}_{\perp}&= \frac{m\sum_{i=1}^N\langle\dot{x}_i^2+\dot{y}_i^2\rangle}{2Nk_b},\label{eq:temp_pl}\\
T^{PE}&= \frac{2}{3}\frac{\sum_{i=1}^N(\phi_r(\boldsymbol{x}_i)-\phi_r(\boldsymbol{x}_i^0))}{Nk_b}.\label{eq:temp_pe}
\end{align}
\end{subequations}

Here, the coordinates are taken to be in the rotating frame, although we have dropped the $r$ subscript.  We use SI units in these calculations and convert the temperatures to units of mK before plotting. It is also important to characterize the ion position fluctuations after laser cooling, as they contribute to the unwanted broadening of the axial mode spectrum, which must be well resolved to implement certain quantum information protocols.  In Figure 5b, we plot a histogram of each ion's 3D root mean square displacement, $\Delta r_{rms}^i$, after laser cooling has been turned off and the crystal has been allowed to evolve freely. This is given by

\begin{equation}
\label{eq_rms_disp}
    \Delta r_{rms}^i = \sqrt{\langle |\boldsymbol{x}_i-\langle\boldsymbol{x}_i\rangle|^2\rangle}.
\end{equation}

Here, the angle brackets represent time-averages over 1 millisecond after the lasers are turned off.  For the particular set of laser parameters used in figure \ref{fig:cooling_single_sim}, it is shown that the rms displacement of each ion is $<0.6\;\mu$m, much smaller than the approximate interparticle spacing of $\zeta\sim 10\;\mu$m.  Here, we roughly estimate the interparticle spacing $\zeta$ by assuming that the crystal has a uniform density and dividing the volume $V$ of the crystal by $N=1000$ such that $V/N = (4/3)\pi\zeta^3$.

In order to characterize laser cooling results for various values of laser beam parameters, we run a scan in which we initialize the same crystal, but vary $w_y$ and $\Delta_{\perp}$.  In figure \ref{fig:cooling_results} we plot the post-laser-cooling temperatures, corresponding to both the kinetic and potential energies.  The planar kinetic energy, which is most sensitive to the laser parameters considered here, can be cooled to $<2$ mK using appropriate parameters (see figure \ref{fig:cooling_results_a}).  We compare the simulated and theoretical $KE_{\perp}$ results in figure 6a and 6b, respectively.  The theoretical results are obtained by extending the methods of \citep{torrisi2016perpendicular} to 3D crystals, and a complete discussion of this calculation is provided in appendix B.  We find that the simulation and theory predict very similar planar temperatures.  The axial kinetic energy is plotted in figure \ref{fig:cooling_results_c} and is cooled to $<1$ mK for each set of planar beam parameters used in this scan. 

The temperature corresponding to the crystal's potential energy is plotted in figure \ref{fig:cooling_results_d}.  Interestingly, for certain laser parameters, the crystal's potential energy after laser cooling is actually lower than that of the initial equilibrium configuration.  This is possible because the equilibrium configuration corresponds to a local, not global, energy minimum.  Therefore, equation \eqref{eq:temp_pe} would lead to negative temperatures if we used the $\boldsymbol{x}_i^0$'s corresponding to the initial equilibrium configuration.  To avoid negative temperatures, we perform another minimization. First, we find the $(w_y, \Delta_{\perp})$ pair which corresponds to the lowest potential energy crystal configuration after laser cooling (see figure~\ref{fig:cooling_results_d}).  Next, we use that configuration for the seed ion positions in another scipy BFGS minimization of the potential energy.  The ion coordinates from this minimization then define the $\boldsymbol{x}_i^0$'s which are used in equation \ref{eq:temp_pe}. This ensures that only positive temperatures are returned.  In the future, the temperature estimates for the potential energy can be improved by finding the local minima separately for every $(w_y, \Delta_{\perp})$ pair.  

As illustrated in figure \ref{fig:cooling_results_d}, the potential energy cools to $\sim 1$ mK for a large range of laser beam parameters. It seems easier to obtain low potential energies using spherical 3D crystals compared to planar ones.  This could be due to improved direct laser cooling of the ExB modes because these modes have an axial component in 3D crystals.  Also, from figure \ref{fig:mode_spec}, we expect that the $\boldsymbol{E}\times\boldsymbol{B}$ and axial mode bandwidths for spherical crystal simulations are not gapped, and therefore the coupling between these modes could be strong \citep{johnson2023rapid}.

\section{\label{sec:conclusion}Conclusion}

Our numerical code efficiently simulates the time evolution of large ion crystals in a Penning trap.  This is an important capability as ion crystals with large $N$ offer key advantages in a variety of experiments spanning fields like quantum information and high energy physics.  Our numerical laser cooling simulations suggest that current experimental approaches used to cool smaller planar crystals can be extended in a straightforward manner to 3D ion crystals containing thousands of ions.  This cooling is an important prerequisite for extending quantum protocols to 3D crystals.  For instance, recent studies have considered prospects for quantum information processing in bilayer crystals, which are a subclass of 3D ion crystals formed in anharmonic trapping potentials \citep{Hawaldar:2023czr}. Efficient laser cooling, including sub-Doppler cooling, was identified as an important requirement for high-fidelity experiments using these crystals. Future studies can aim to determine if the laser cooling of 3D crystals demonstrated here can be adapted to such crystals, which will serve as the starting point to explore pathways for sub-Doppler cooling. 

It appears significantly easier to cool the low frequency $\boldsymbol{E}\times\boldsymbol{B}$ modes in 3D crystals than in planar crystals.  A couple of mechanisms are likely responsible for this, and future work will seek to characterize their behavior more precisely.  First, since the frequency gap between mode branches is smaller for 3D crystals than for planar ones, resonant coupling may be partially responsible for cooling, as seen in \cite{johnson2023rapid}.  Furthermore, $\boldsymbol{E}\times\boldsymbol{B}$ and cyclotron modes, which have historically been difficult to cool in planar crystals, gain an axial component in 3D crystals.  Since axial motion is generally easier to cool, the cooling of these modes may be further enhanced in 3D crystals.

Finally, the FMM-enhancement introduced in this paper can be utilized to accelerate different types of ion crystal simulations.  For instance, we plan to investigate the use of artificial damping to calculate crystal equilibria, wherein a nonequilibrium crystal is initialized and allowed to evolve according to the Penning trap equations of motion, while the momenta of the ions are slowly decreased.  This method will be faster than the current SciPy minimization for large ion numbers and, while this method has been used in previous studies, it will be accelerated by a factor of $N$ via the FMM.

\section*{Acknowledgements}
The authors thank Stefan Tirkas for help with programming the Penning trap simulation. We thank Joseph Andress and Allison Carter for reading the manuscript and providing helpful feedback.  We appreciate our insightful discussions with Bryce Bullock, Allison Carter, Dan Dubin, Samarth Hawaldar, Leslie Greengard, and Jennifer Lilieholm. This work was supported by the U.S. Department of Energy under Grant No. DE-SC0020393.  J.J.B. acknowledges support from DOE, Office of Science, Quantum Systems Accelerator, from AFOSR, and from the DARPA ONISQ program. A.S. acknowledges the support of a C.V. Raman Post-Doctoral Fellowship, IISc.

\appendix

\section{Laser Cooling Theory}\label{appB}

Theoretical treatments of the laser cooling of ion crystals in a Penning trap have been used to estimate the minimum achievable temperatures corresponding to $KE_{\perp}$ \citep{itano1988perpendicular,torrisi2016perpendicular}.  However, to our knowledge, the temperatures of 3D crystals subject to a rotating wall potential have not been estimated.  Here, we apply previously developed theoretical techniques to study this case.  

We analyze the scattering of photons from the planar beam by the ion crystal in order to predict the planar temperature, $T_{\perp}$, after laser cooling. The photon scattering rate of a single ion with mass $m$, position $(x,y,z)$ and velocity $(v_x,v_y,v_z)$ in a laser field with frequency $\omega_L$ and intensity $I_{\perp}(y,z)$ which propagates in the $\hat{x}$ direction is given by

\begin{equation}
    \gamma_L(y,z, v_x) = \frac{I_{\perp}(y,z)\sigma_0}{\hbar\omega_L}\frac{(\gamma_0/2)^2}{[\gamma_{\perp}(y,z)/2]^2 + (\Delta_{\perp}-kv_x)^2}.
\end{equation}

Here, $\gamma_0$ and $\sigma_0$ are the natural linewidth and scattering cross-section on resonance, respectively, of the $2s^2S_{1/2}\rightarrow 1p^2P_{3/2}$ cooling transition. $\Delta_{\perp}$ is the detuning of the laser from the transition frequency, and $k$ is the wavenumber given approximately by $k=\omega_0/c$.  For a 2D Gaussian beam propagating in the $\hat{x}$ direction and centered on $(y,z) = (d,0)$, we have $I_{\perp}(y,z) = I_{\perp,0}\exp[-\frac{(y-d)^2}{w_y^2}-\frac{z^2}{w_z^2}]$.  In terms of the saturation intensity $S_{\perp}(y,z) = I_{\perp}(y,z)\sigma_0/\hbar\omega_0\gamma_0$, the saturation adjusted linewidth is $\gamma_{\perp}(y,z) = \gamma_0\sqrt{1+2S_{\perp}(y,z)}$.  The scattering rate is then approximately given by

\begin{equation}
    \gamma_L(y,z, v_x) = \frac{\gamma_0 S_{\perp,0} \exp[-\frac{(y-d)^2}{w_y^2}-\frac{z^2}{w_z^2}]}{1 + 2S_{\perp,0} \exp[-\frac{(y-d)^2}{w_y^2}-\frac{z^2}{w_z^2}] +\big(\frac{2}{\gamma_0}\big)^2(\Delta_{\perp}-kv_x)^2}.
\end{equation}

Here, $S_{\perp,0} = I_{\perp,0}\sigma_0/\hbar\omega_0\gamma_0$ is the saturation parameter, defined as the maximum saturation intensity. The rate per unit volume at which the ion crystal scatters photons is expressed as

\begin{equation}
    S(x,y,z)  = \int_{-\infty}^{\infty}dv_xP(v_x|y,u)\rho(x,y,z)\gamma_L(y,z, v_x) .
\end{equation}

Here, $\rho$ is the three-dimensional number density of ions in the crystal. $P(v_x|y,u)$ is the probability distribution of $v_x$ in terms of $u = \sqrt{2k_BT_{\perp}/m}$. It is described by a Maxwell-Boltzmann velocity distribution in the frame rotating at frequency $\omega_r$ which, in the lab frame, is written as 

\begin{equation}
    P(v_x|y,u)  = \frac{1}{u\sqrt{\pi}}\exp\Big[\frac{-(v_x-\omega_ry)^2}{u^2}\Big].
\end{equation}

The time-averaged cooling rate of the planar kinetic energy due to the laser is obtained by multiplying the integrand of $S$ by the average change in the planar kinetic energy due to a single photon scattering event and then integrating over $v_x$ as well as the three spatial coordinates of the crystal.  This assumes no coupling between planar and axial degrees of freedom.  The cooling rate is given by

\begin{equation}
    \Big\langle\frac{dE}{dt}\Big\rangle_{laser}  = \int dz\int dy\int dx\int_{-\infty}^{\infty}dv_x\Big(\hbar k v_x+\frac{5}{3}\frac{\hbar^2k^2}{2m}\Big)P(v_x|y,u)\rho(x,y,z)\gamma_L(y,z, v_x) .
\end{equation}

Since the rotating wall potential does work on the ion crystal to maintain a constant rotation frequency, this change in energy must also be considered.  To find the rate of energy change due to the rotating wall potential, we assume that the only external torques on the ion crystal come from the rotating wall and the planar cooling laser such that, in equilibrium, $\tau_{wall} = -\tau_{laser}$.  This implies

\begin{equation}
\begin{split}
    \Big\langle\frac{dE}{dt}\Big\rangle_{wall}  = -\omega_r \int dz\int dy\int dx\hbar kyS(x,y,z) .
\end{split}
\end{equation}

The total rate of change of the ion crystal's energy due to the planar beam and the rotating wall is then

\begin{equation}
\begin{split}
    \Big\langle\frac{dE}{dt}\Big\rangle  = \int dz\int dy\int dx\int_{-\infty}^{\infty}dv_x\Big(\hbar k v_x+\frac{5}{3}\frac{\hbar^2k^2}{2m} - \omega_r\hbar k y\Big)P(v_x|y,u)\rho(x,y,z)\gamma_L(y,z, v_x)  .
\end{split}
\end{equation}

Setting $v = (v_x-\omega_ry)/u$, $v_{rec} = 5\hbar k/ 6m$, $\delta_y = (y-d)/w_y$, and $\delta_z = z/w_z$, and taking $\rho$ to be uniform over the volume of the crystal, the above expression is expressed as

\begin{equation}
\label{eq: total_dedt}
    \scalebox{1.05}{$
    \Big\langle\frac{dE}{dt}\Big\rangle 
    =\frac{\gamma_0 S_{\perp,0}\rho\hbar k}{\sqrt{\pi}}u\int dz\int dy\int dx\int_{-\infty}^{\infty}dv\frac{\big(v+\frac{v_{rec}}{u}\big)\exp(-v^2)  \exp[-\delta_y^2-\delta_z^2]}{1 + 2S_0 \exp[-\delta_y^2-\delta_z^2] +\big[\frac{\Delta_{\perp}-k\omega_rd}{\gamma_0/2}-\frac{k\omega_rw_y\delta_y}{\gamma_0/2}-\frac{kv_{rec}}{\gamma_0/2}\frac{u}{v_{rec}}v\big]^2}$}.
\end{equation}

The integral over $x$ is trivial since the integrand does not depend on $x$.  If only the planar beam is used to cool the crystal, then one can solve for the post-cooling planar temperature $T_{\perp}$ by solving for $u_{eq}$ such that $\langle dE/dt\rangle(u_{eq}) =0$.  As described in the main text, however, axial beams are also used in our simulations to cool the the motion of the ions along the $\hat{z}$ direction.  These beams, characterized by saturation parameter $S_{\parallel,0}$, cause recoil heating in the planar directions, leading to temperatures slightly greater than predicted by equation \ref{eq: total_dedt}.  To account for this, we add to equation \ref{eq: total_dedt} the following heating term describing the planar recoil heating due to a single axial beam.

\begin{equation}
\begin{split}
    \Big\langle\frac{dE}{dt}\Big\rangle_{recoil}  = \frac{\gamma_0S_{\parallel,0}N}{3(1+S_{\parallel,0})} \frac{(\hbar k)^2}{2m}
\end{split}
\end{equation}

Here, $N$ is the total number of ions in the crystal. This expression is valid when the intensity of each axial beam is taken to be uniform throughout the crystal.  The theoretical results shown in figure \ref{fig:cooling_results_b} include the recoil heating contributions and agree with the simulation results in figure \ref{fig:cooling_results_a}.

\section{Ion Crystal Normal Modes}\label{appA}
In this appendix we derive the equations of motion of a general $N$-ion 3D ion crystal in a Penning trap with a rotating wall potential and explain how the normal modes are obtained. In the frame rotating at frequency $\omega_r$, it can be shown that the Lagrangian of the ion crystal system is given by 

\begin{equation}
\label{app_lagrangian}
\begin{split}
    \mathcal{L} = \sum_{j=1}^N\Bigg[&\frac{1}{2}m|\dot{\boldsymbol{x}}_j|^2 - \frac{q}{2}B_{eff}(\dot{x}_jy_j-x_j\dot{y}_j)-\frac{1}{2}m\omega_z^2z_j^2\\
    &+\frac{1}{2}m\Big(\omega_r^2-\omega_c\omega_r+\frac{1}{2}\omega_z^2\Big)(x_j^2+y_j^2)-\frac{1}{4}qk_z\delta(x_j^2-y_j^2) -\frac{kq^2}{2}\sum_{i\neq j}\frac{1}{|\boldsymbol{x}_i-\boldsymbol{x}_j|}\Bigg].
\end{split}
\end{equation}

Here, $B_{eff} = (B-2m\omega_r/q)$ is the effective magnetic field which the ions see in the rotating frame. We will denote the ions' equilibrium positions in the rotating frame using $\{\boldsymbol{x}^0_i\}$.  Near equilibrium, the ion positions and velocities are denoted $\{\boldsymbol{\delta x}_i\}$ and $\{\boldsymbol{v}_i\}$, respectively. The dynamics near equilibrium are found by expanding the Lagrangian about the crystal's equilibrium configuration.

\begin{equation}
\label{lagrangian}
\begin{split}
   L &=  L_{eq}+\frac{1}{2}\sum_{ij}\sum_{\alpha\beta}\Big(\frac{\partial^2 L}{\partial \alpha_i\partial \beta_j}\Big|_{\{\boldsymbol{x}^0_i\}}\delta \alpha_i\delta \beta_j + \frac{\partial^2 L}{\partial \dot{\alpha}_i\partial \dot{\beta}_j}\Big|_{\{\boldsymbol{x}^0_i\}}\delta \dot{\alpha}_i\delta \dot{\beta}_j+2\frac{\partial^2 L}{\partial \dot{\alpha}_i\partial \beta_j}\Big|_{\{\boldsymbol{x}^0_i\}}\delta \dot{\alpha}_i\delta \beta_j\Big)\\
   &= L_{eq}-\frac{1}{2}\sum_{ij}\sum_{\alpha\beta}K_{ij}^{\alpha\beta}\delta \alpha_i\delta \beta_j + \frac{1}{2}\sum_{ij}\sum_{\alpha\beta}M_{ij}^{\alpha\beta}\delta \dot{\alpha}_i\delta \dot{\beta}_j+\sum_{ij}\sum_{\alpha\beta}W_{ij}^{\alpha\beta}\delta \dot{\alpha}_i\delta \beta_j\\
\end{split}
\end{equation}

Here, $\alpha,\beta\in \{x,y,z\}$ and $i,j\in \{1,2,\cdots,N\}$. It is then straightforward to derive the Euler-Lagrange equations of motion, which are conveniently to written in terms of a length $6N$ vector $\ket{q}$ given by

\begin{equation}
\begin{split}
\label{app_q}
    \boldsymbol{q} =\big(
    \delta x_1 \dots\delta {x}_N&\delta{y}_1  \dots \delta{y}_N\delta{z}_1  \dots \delta{z}_N\dot{x}_1\dots\dot{x}_N \dot{y}_1\dots\dot{y}_N\dot{z}_1\dots\dot{z}_N
    \big)^T.
\end{split}
\end{equation}

The time evolution of $\boldsymbol{q}$ is then described by

\begin{equation}
\label{app_eom}
\begin{split}
    \frac{d\boldsymbol{q}}{dt}= 
    \begin{pmatrix}
    \mathsfbi{0}&\mathsfbi{0}&\mathsfbi{0}&\mathsfbi{1}&\mathsfbi{0}&\mathsfbi{0}\\\mathsfbi{0}&\mathsfbi{0}&\mathsfbi{0}&\mathsfbi{0}&\mathsfbi{1}&\mathsfbi{0}\\\mathsfbi{0}&\mathsfbi{0}&\mathsfbi{0}&\mathsfbi{0}&\mathsfbi{0}&\mathsfbi{1}\\
    -\mathsfbi{K}^{xx}/m&-\mathsfbi{K}^{xy}/m &-\mathsfbi{K}^{xz}/m  &\mathsfbi{0}&-2\mathsfbi{W}^{xy}/m&\mathsfbi{0}\\-\mathsfbi{K}^{xy}/m& -\mathsfbi{K}^{yy}/m&-\mathsfbi{K}^{yz}/m  &-2\mathsfbi{W}^{yx}/m&\mathsfbi{0}&\mathsfbi{0}\\-\mathsfbi{K}^{xz}/m& -\mathsfbi{K}^{yz}/m&-\mathsfbi{K}^{zz}/m  &\mathsfbi{0}&\mathsfbi{0}&\mathsfbi{0}
    \end{pmatrix}\boldsymbol{q}.
\end{split}
\end{equation}

Defining $r^0_{ij} = |\boldsymbol{x}_i^0-\boldsymbol{x}_j^0|$ and $\alpha^0_{ij} = |\alpha_i^0-\alpha_j^0|$ (for $\alpha\in\{x,y,z\}$), the various $\mathsfbi{W}^{\alpha\beta}$ and $\mathsfbi{K}^{\alpha\beta}$ are explicitly given by (in terms of $k_z$ and $B$ rather than $\omega_z$ and $\omega_c$)  

\begin{equation}
\label{W_matrix}
    W^{xy}_{ij} =-W^{yx}_{ij}=\frac{qB_{eff}}{2}\delta_{ij},
\end{equation}

\begin{equation}
\label{Kxx}
\begin{split}
    K^{xx}_{ij} =\begin{cases}
m\omega_r^2+qB_{eff}\omega_r-\frac{qk_z}{2}+\frac{qk_z\delta}{2}-kq^2\sum_{k\neq i} \frac{r_{ik}^{0^2}-3x^{0^2}_{ik}}{r_{ik}^{0^5}} &i=j\\
kq^2\frac{r_{ij}^{0^2}-3x^{0^2}_{ij}}{r_{ij}^{0^5}} &i\neq j ,\\
\end{cases}
\end{split}
\end{equation}

\begin{equation}
\label{Kyy}
\begin{split}
    K^{yy}_{ij} =\begin{cases}
m\omega_r^2+qB_{eff}\omega_r-\frac{qk_z}{2}-\frac{qk_z\delta}{2}-kq^2\sum_{k\neq i} \frac{r_{ik}^{0^2}-3y^{0^2}_{ik}}{r_{ik}^{0^5}} &i=j\\
kq^2\frac{r_{ij}^{0^2}-3y^{0^2}_{ij}}{r_{ij}^{0^5}} &i\neq j ,\\
\end{cases}
\end{split}
\end{equation}

\begin{equation}
\label{Kzz}
\begin{split}
    K^{zz}_{ij} =\begin{cases}
m\omega_z^2-kq^2\sum_{k\neq i} \frac{r_{ik}^{0^2}-3z^{0^2}_{ik}}{r_{ik}^{0^5}} &i=j\\
kq^2\frac{r_{ij}^{0^2}-3z^{0^2}_{ij}}{r_{ij}^{0^5}} &i\neq j ,\\
\end{cases}
\end{split}
\end{equation}

\begin{equation}
\label{Kab}
\begin{split}
    K^{\alpha\beta}_{ij} =\begin{cases}
kq^2 \sum_{k\neq i}\frac{3\alpha_{ik}^0\beta_{ik}^0}{r_{ik}^{0^5}} &i=j, \alpha\neq\beta\\
-kq^2\frac{3\alpha_{ij}^0\beta_{ij}^0}{r_{ij}^{0^5}} &i\neq j,\alpha\neq\beta .\\
\end{cases}
\end{split}
\end{equation}

The equation \ref{app_eom} is very similar to the analogous equations of motion for a planar crystal.  However, for a planar crystal, $K_{ij}^{xz} = K_{ij}^{yz} = 0$ for all $\{i,j\}$. The eigenmodes of the ion crystal system are obtained by diagonalizing the above matrix.


\bibliographystyle{jpp}

\bibliography{jpp-instructions}

\end{document}